\documentclass[aps,pra,superscriptaddress,twocolumn]{revtex4-2}

\usepackage{amsmath} 
\usepackage{graphicx}
\usepackage{mathtools}
\usepackage[table,svgnames]{xcolor}
\usepackage{subfigure}
\usepackage{makecell}
\usepackage{hyperref}
\hypersetup{colorlinks=true,        
    linkcolor=black,         
    citecolor=black,        
    filecolor=black,      
    urlcolor=black }

\usepackage{booktabs}

\newcommand{\pd}[2]{\frac{\partial #1}{\partial #2}} 
\newcommand{\ket}[1]{\left| #1 \right>} 
\newcommand{\bra}[1]{\left< #1 \right|} 

\newcommand{\tom}[1]{{\color{brown}{#1}}}

\newcommand{\abs}[1]{| #1 |} 
\newcommand{\absa}[1]{\left| #1 \right|} 
\newcommand{\sech}[1]{\text{sech}\left(#1\right)} 
\newcommand{\pt}{T} 
\newcommand{\ct}{\sigma_{t}} 

\begin{document}


\title{
Fast storage of photons in cavity-assisted quantum memories
}


\author{Johann S. Kollath-Bönig}
\affiliation{Niels Bohr Institute, University of Copenhagen, Blegdamsvej 17, 2100 Copenhagen Ø, Denmark}
\affiliation{Center for Hybrid Quantum Networks, Niels Bohr Institute,
University of Copenhagen, Blegdamsvej 17, 2100, Copenhagen, Denmark}
\author{Luca Dellantonio}
\affiliation{Department of Physics and Astronomy, University of Exeter, Stocker Road, Exeter EX4 4QL, United Kingdom}
%
\author{Luigi Giannelli}
\affiliation{Dipartimento di Fisica e Astronomia ``Ettore Majorana'', Università di Catania, Via S. Sofia 64, 95123 Catania, Italy}
\affiliation{CNR-IMM, UoS Università, 95123 Catania, Italy}
\affiliation{INFN Sezione di Catania, 95123 Catania, Italy}
%
\author{Tom Schmit}
\affiliation{Theoretische Physik, Saarland University, 66123 Saarbr\"ucken, Germany}
\author{Giovanna Morigi}
\affiliation{Theoretische Physik, Saarland University, 66123 Saarbr\"ucken, Germany}
%
\author{Anders S. Sørensen }
\affiliation{Niels Bohr Institute, University of Copenhagen, Blegdamsvej 17, 2100 Copenhagen Ø, Denmark}
\affiliation{Center for Hybrid Quantum Networks, Niels Bohr Institute,
University of Copenhagen, Blegdamsvej 17, 2100, Copenhagen, Denmark}


\date{\today}

\begin{abstract}

Ideal photonic quantum memories can store arbitrary pulses of light with unit efficiency. This requires operating in the adiabatic regime, where pulses have a duration much longer than the bandwidth of the memory. In the nonadiabatic regime of short pulses, memories are therefore imperfect, and information is always lost. We theoretically investigate the bandwidth limitations for setups based on individual atoms, or ensembles thereof, confined inside optical cavities. We identify an effective strategy for optimizing the efficiencies of the storage and retrieval process regardless of the duration or the shape of the pulses. Our protocol is derived almost completely analytically and attains efficiencies that are comparable to those obtained by numerical optimization. Furthermore, our results provide an improved  understanding of the performance of quantum memories in several regimes. 
When considering pulses defined on an infinite time interval, the shapes can be divided into two categories, depending on their asymptotic behaviors. 
If the intensity of the pulse increases with time slower than or as an exponential function, then the storage efficiency is only limited by the pulse width.
For pulses defined on a finite interval, on the other hand, the efficiency is determined by the shape at the beginning of the storage or, correspondingly, at the end of the retrieval process.
\end{abstract}


\maketitle

\section{Introduction}
\label{sec:Intro}
Quantum memories that can store and retrieve incoming light pulses are an essential building block for quantum information technologies, as they allow local storage and processing of optical signals carrying quantum information between different nodes.
As a particular example, some of the most promising quantum repeater schemes for long-distance quantum communication resort to quantum memories for storing light pulses \cite{Duan2001,Simon2007,Sangouard2011Quantum,simon2010quantum,Fleischhauer2002Quantum}.
Therefore, several proposals and experiments have explored methods to create quantum memories based on either individual \cite{Cirac1997QuantumState,Dilley2012Single,Tatjana2007Single,specht2011single,McKeever2004Deterministic,Tatjana2007Single,Giannelli2018,Giannelli2019Weak,Kurz2016Programmable,Mundt2002Coupling,Boozer2007Reversible,Tolazzi2021} or ensembles of atoms \cite{Julsgaard2004,Kraus2006Quantum,Afzelius2009Multimode,Hammerer2010Quantum,choi2008mapping,cao2020efficient,nicolas2014quantum,Gouraud2015Demonstration,Laurat2006Atomic,Veselkova2019,zhao2009millisecond,himsworth2011eit,Gorshkov2007Universal}.  

For a wide class of these memories, 
it was shown in Refs.~\cite{Gorshkov2007Universal,Gorshkov2007a,Gorshkov2007b,Fleischhauer2002Quantum,Zhang2023Limits,Dilley2012Single,Cirac1997QuantumState,himsworth2011eit} that in the long pulse limit it is always possible to perfectly store and retrieve the desired quantum information.
In practice, however, it is often desirable to work with short pulses. For instance, single-photon sources, e.g. based on quantum dots \cite{Lodahl2015Interfacing,Arakawa2020Progress,Li2023Quantum,Maentynen2019Single}, are typically broadband, whereas atomic memories have a much smaller bandwidth due to their longer lifetimes \cite{Julsgaard2004,Tolazzi2021,Boozer2007Reversible,Trotzky2010Controlling}. In this regime, photon memories always have limited efficiencies both in storing and retrieving quantum information.
It is therefore essential to ensure their optimal performance given the available experimental resources. 
%
Multiple studies have investigated methods to attain this optimal performance in the nonadiabatic regime. Several were based on analytical (see, e.g., Refs.~\cite{Gorshkov2007Universal,Afzelius2010Impedance}) or numerical (see, e.g., Refs.~\cite{Giannelli2018,Macha2020,Cai2021Optimizing}) maximization of the efficiency.
To date, however, an optimal storage and retrieval strategy in the nonadiabatic regime has not been identified.

In this article, we analyze the theoretical limits to the storage and retrieval efficiencies in the fast, nonadiabatic (i.e., short-pulse) regime. We consider quantum memories in optical cavities and explicitly evaluate the relation between the arbitrary pulse shape of the incoming to-be-stored (or outgoing to-be-retrieved) photon and the efficiency of the memory. The developed theory applies equally to memories based on atomic ensembles and individual atoms, provided that the incoming light fields are sufficiently weak, i.e. contain much fewer photons than the number of atoms in the ensemble or at most a single photon in the case of a single atom.

To better understand the memory response and performance, we analyze two complementary regimes determined by the characteristic time scales of cavity and atoms: one that we call the ``atom-limited" regime, where the characteristic timescale of the photon pulse is much longer than the characteristic timescale of the cavity, such that the bandwidth is limited by the atomic response; and one that we call the ``cavity-limited'' regime, where the roles of the cavity and atomic ensemble are reversed.
In these two different limits, we identify a universal strategy for optimizing the memory efficiency. Interestingly, we find essentially the same behavior, provided we exchange the atomic (for the ``atom-limited'' regime) and cavity (for the ``cavity-limited'' regime) response times.

Our approach to the optimization is almost completely analytical, and
is based on an ansatz with a single free parameter, which may need to be optimized numerically. Despite its simplicity, our strategy performs better than or equal to full numerical optimizations that generally require much larger computational resources. Our approach further allows a detailed understanding of how the shape of the pulses affect the storage and retrieval efficiencies. In particular, we find that the storage (retrieval) efficiency critically depends on the initial and final stages of the dynamics, when the memory interacts with the pulse tails. Related optimization strategies have been considered in Refs.~\cite{Vasilev2010,Utsugi2022,Tissot2024}, where a different ansatz for the ideal shape is employed. As we argue below, both ours and these strategies reach unit efficiency for sufficiently long pulses. When the efficiency deviates from unity, however, we show that our approach gives higher efficiencies. 



The manuscript  is structured as follows. In Sec.~\ref{sec:Model}, we describe the considered setup and introduce the 
figure of merit characterizing the quality of the memory. The ansatz we employ for achieving optimality is outlined in Sec.~\ref{sec:Bad_cavity_limit} for the ``atom-limited'' regime, which is attained for pulse lengths much longer than the cavity lifetime. Importantly, we consider both truncated and infinitely long pulses, and identify the features that are most detrimental to the storage and retrieval efficiencies. In Sec.~\ref{sec:Good_cavity_limit} we expand the results to the opposite ``cavity-limited'' regime, which is compared with the ``atom-limited'' regime in Sec.~\ref{sec:connection_regimes}. Finally, conclusions are presented in Sec.~\ref{sec:conclusions}. In the appendices we report analytical and numerical details on the derivation of the optimal protocol and on its efficiency for specific pulse shapes.

\section{Model}
\label{sec:Model}
\begin{figure}[]
	\centering
	\includegraphics[width=\columnwidth]{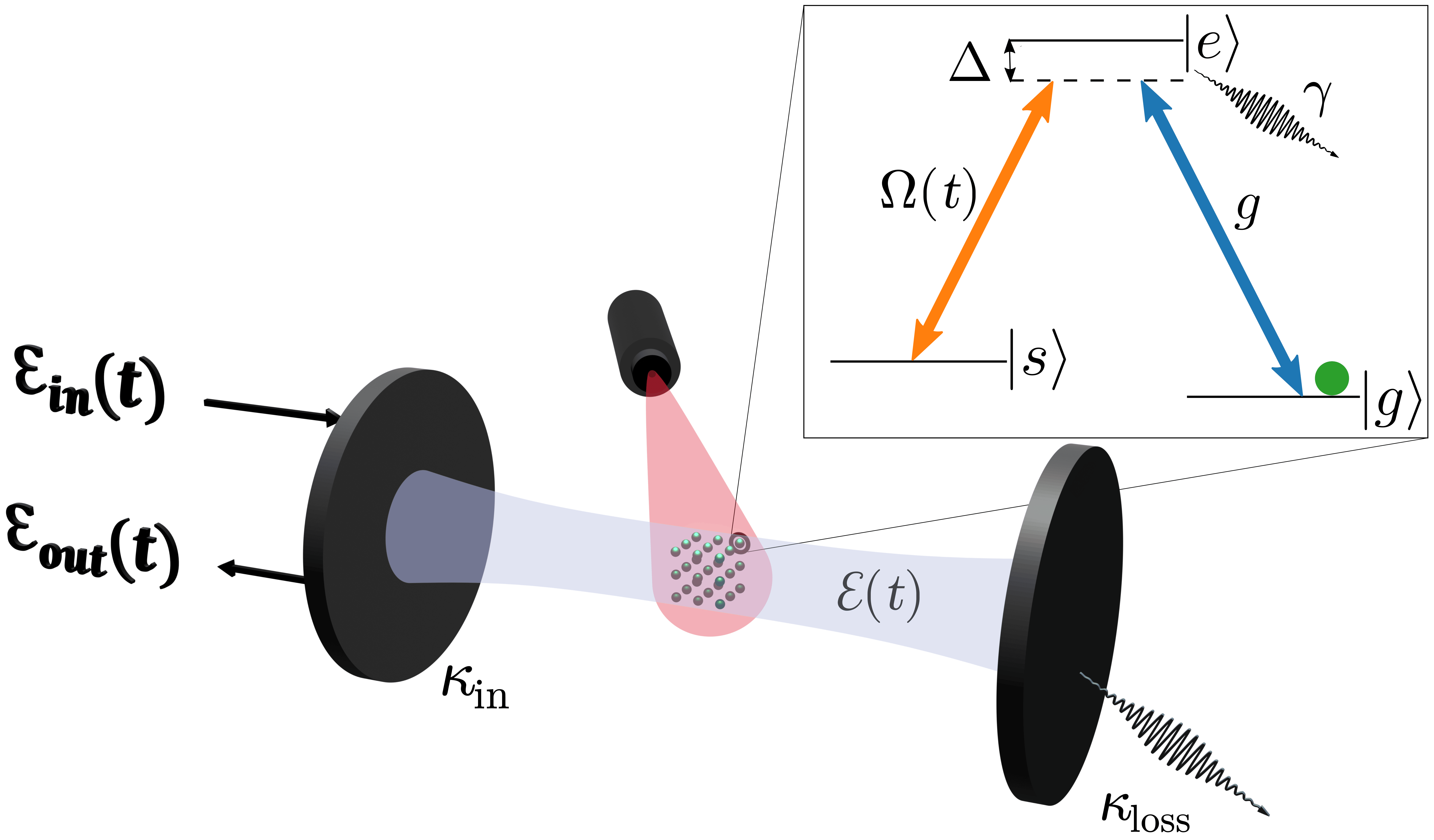}
	\caption{
 Schematic representation of the considered quantum memory. A single atom or an ensemble of atoms is trapped inside an optical cavity with incoupling and intrinsic decay rates $\kappa_\text{in}$ and $\kappa_\text{loss}$, respectively. The atoms are modeled as three-level systems and are initially prepared in $\ket{g}$ (green dot). The quantized cavity field (blue) couples with strength $g$ to the transition $\ket{g}$-$\ket{e}$, allowing the transfer of a cavity photon to an atomic excitation. The control field $\Omega(t)$ (orange) coupling levels $\ket{e}$ and $\ket{s}$ allow for the transfer of such excitations to the stable storage level $\ket{s}$, enabling a long-term quantum memory. 
 Both control $\Omega(t)$ and incoming $\mathcal{E}_{{\rm in}}(t)$ fields are characterized by a detuning $\Delta$ with respect to the associated transitions and the excited state $\ket{e}$ decays at a rate $\gamma$.}
	\label{fig:raman}
\end{figure}

As schematically represented in Fig.~\ref{fig:raman}, we consider a quantum memory made of $N$ $\Lambda$-type atoms \cite{marangos1998electromagnetically} inside a cavity. The storage and retrieval processes are based on Raman transitions between the two stable electronic states $\ket{g}$ and $ \ket{s}$ via the excited level $\ket{e}$. The annihilation operator associated with the optical field inside the cavity is $\hat{\mathcal{E}}$, which is coupled to the incoming and outgoing fields $\hat{\mathcal{E}}_\text{in}$ and $\hat{\mathcal{E}}_\text{out}$, respectively, via a partially transparent mirror (to the left) with decay rate $\kappa_\text{in}$. For convenience, we assume that the right mirror is totally reflective.
During the storage process, the incoming photon field $\hat{\mathcal{E}}_\text{in}$ couples to the cavity field $\hat{\mathcal{E}}$, which in turn drives the electronic transition $\ket{g}\to\ket{e}$ of the atoms
initially prepared in $\ket{g}$. 
The atom-cavity coupling is characterized by the vacuum Rabi frequency $g$.
By applying a control field with amplitude $\Omega(t)$ to the $\ket{s}$-$\ket{e}$ transition, the atomic population is
transferred to $\ket{s}$, thus completing the storage process.

In this work, we assume that during storage a single photon or a weak quantum field is sent towards the cavity. This photon is in a single optical mode with a pulse shape $\phi_s(t)$, such that it can be described by the single-mode photon annihilation operator $\hat{\phi}_s$, defined as: 
\begin{equation}\label{eq:phi_s}
    \hat{\phi}_s = \int_{-\infty}^{\infty} \phi_{s}^{*}(t) \hat{\mathcal{E}}_\text{in}(t) dt\,,
\end{equation}
where $t$ indicates the time at which the field arrives at the cavity interface and $\phi_s(t)$ denotes the temporal behavior of the pulse amplitude.
Normalization of $\phi_s$, i.e., $\int \phi_s^*(t')\phi_s(t')dt' = 1$ ensures that the single-mode operator $\hat{\phi}_s$ is well defined and respects standard commutation relations with its adjoint $[\hat{\phi}_s,\hat{\phi}_{s}^{\dagger}] = 1$. 

When one or more atoms are driven into $\ket{s}$, we say that an excitation has been transferred from $\ket{g}$ and stored into $\ket{s}$ (mediated by the fields $\hat{\mathcal{E}}$ and $\Omega$). We assume that the incoming and control fields are at two-photon resonance with detuning $\Delta$ with respect to the excited state $\ket{e}$. Herewith we mean that their carrier frequencies are detuned by the same amount $\Delta$ from the transitions that they drive ($\ket{g}$-$\ket{e}$ and $\ket{s}$-$\ket{e}$, respectively); see Fig.~\ref{fig:raman}. During retrieval the process is reversed such that the driving field  $\Omega$ excites the atoms from $\ket{s}$ to $\ket{e}$, thus emitting a photon into the cavity that subsequently leaks into the output field $\hat{\mathcal{E}}_\text{out}$. As explained in detail below, our goal is not only to recover the state into a general field $\hat{\mathcal{E}}_\text{out}$, but to have its shape as close as possible to a specific mode $\phi_r$. 
Losses in the system are incorporated by the decay of the optical coherence at a rate $\gamma$ and at the additional cavity field decay rate $\kappa_\text{loss}$.


For describing the system's dynamics, we employ the Jaynes–Cummings model \cite{larson2022jaynes}, yielding the following Hamiltonian in the dipole and rotating-wave approximations:
\begin{equation}\label{eq:sys_ham}
    \hat{H}=\hbar\Delta\hat{\sigma}_{ee}-\left(\hbar\Omega(t)\hat{\sigma}_{es}+\hbar \hat{\mathcal{E}}g\hat{\sigma}_{eg}+\text{H.c.}\right).
\end{equation}
Here,
\begin{equation}
    \hat{\sigma}_{\mu\nu}=\sum_{j=1}^N\ket{\mu}_j{\bra{\nu}}_{j}e^{i\phi^{(j)}_{\mu\nu}} \qquad \text{for $\mu,\nu = g,s,e$}
\end{equation}
are collective atomic operators with phases $\phi^{(j)}_{\mu\nu}$, that depend on the 
wave number of the incoming fields and the atomic position(s) and are zero for $\mu = \nu$ (see also Ref.~\cite{Gorshkov2007a}).
%
%
%
%
The analytical description of the system characterized by Hamiltonian~\eqref{eq:sys_ham} is greatly simplified by the fact that the dynamics is  restricted to a few collective atom-light states. If, at the beginning of the storage, all atoms are initialized in
$\ket{g}$ and a single photon is incident, there are only four possible states characterizing the whole dynamics. We denote these states by
\begin{equation}\label{eq:states_dynamics}
    \left\lbrace\ket{G,0,1},\ket{G,1,0},\ket{E,0,0},\ket{S,0,0}\right\rbrace.
\end{equation}
%
The first quantum number describes the states of the atom(s), 
where $\ket{G}$ is the joint ground state of all atoms $\ket{G} = \ket{g}^{\otimes N}$ and $\ket{E}$ ($\ket{S}$) is a superposition of all possible states with one atom in $\ket{e}$ ($\ket{s}$) and all others in $\ket{g}$. The second and third quantum numbers in Eq.~\eqref{eq:states_dynamics} correspond to
the number of cavity photons and the number of photons in the propagating mode, respectively. 
These states form a basis of the reduced Hilbert space in which the dynamics takes place. In the following, for brevity, we indicate with $\ket{g}$, $\ket{e}$, and $\ket{s}$ both the corresponding single atom's levels, and the collective states $\ket{G}$, $\ket{E}$, and $\ket{S}$. Radiative decay of state $\ket{E}$ and photon losses from the cavity imperfections irreversibly couple the subspace of Eq.~\eqref{eq:states_dynamics} to state $\ket{G,0,0}$, where the dynamics stops, while absorption of thermal optical photons from the environment is here neglected.


To describe the cavity system, we use the input-output formalism \cite{gardiner2004quantum,kiilerich2019input} that prescribes 
\begin{equation}
\label{eq:in-out}
\hat{\mathcal{E}}_{\text{out}}(t)=\sqrt{2\kappa_\text{in}}\hat{\mathcal{E}}(t)-\hat{\mathcal{E}}_{\text{in}}(t).
\end{equation}
%
The atomic ensemble stores the incoming field $\hat{\mathcal{E}}_\text{in}$ (and thus $\hat{\phi}_s$) as collective spin-wave excitations. These can be described \cite{Gorshkov2007a} by the approximate annihilation operator $\hat{S}(t)=\hat{\sigma}_{gs}(t)/\sqrt{N}$. Similarly, we define the polarization annihilation operator $\hat{P}(t)=\hat{\sigma}_{ge}(t)/\sqrt{N}$.
We then obtain the Heisenberg-Langevin equations of motion
\begin{subequations}
\label{eq:eqm}
\begin{align}
\dot{\hat{\mathcal{E}}} 
=&
-\kappa\hat{\mathcal{E}}+ig\sqrt{N}\hat{P}+\sqrt{2\kappa_\text{in}}\hat{\mathcal{E}}_\text{in}
+\sqrt{2\kappa_\text{loss}}\hat{\mathcal{E}}_\text{loss},\label{eq:eqm1}\\
\dot{\hat{P}}
=&
-\left(\gamma+i\Delta\right)\hat{P}+ig\sqrt{N}\hat{\mathcal{E}}+i\Omega\hat{S}+\hat{F}_{ge}/\sqrt{N}
,\label{eq:eqm2}\\
\dot{\hat{S}}
=&
i\Omega^*\hat{P}
,\label{eq:eqm3}
\end{align}
\end{subequations}
where $\kappa=\kappa_\text{in}+\kappa_\text{loss}$ is the total cavity decay rate. The noise operators $\hat{F}_{ge}$ and $\hat{\mathcal{E}}_\text{loss}$ in these expressions are the Heisenberg-Langevin operators relative to the associated decay of the $\ket{g}-\ket{e}$ optical coherence
and the cavity mode, respectively \cite{Giannelli2018,Nathan2020Universal}. Note that
we neglect any decoherence of the ground-state coherence $S$, 
assuming that the storage and retrieval occur on timescales that are much shorter than the dephasing or decoherence processes affecting the $\ket{s}$ and $\ket{g}$ levels, i.e., we assume that the system is a good memory. 
Furthermore, since the dynamics is constrained within the states in Eq.~\eqref{eq:states_dynamics}, for deriving the equations of motion Eqs.~\eqref{eq:eqm} we have set $\hat{\mathcal{E}}\hat{\sigma}_{ee}=\hat{\mathcal{E}}\hat{\sigma}_{es}=0$ and $\hat{\mathcal{E}}\hat{\sigma}_{gg}=\hat{\mathcal{E}}N$.

In this work, we analyze and optimize the performance of quantum memories in terms of the storage $\eta_s$ and retrieval $\eta_r$ efficiencies.
As  explained in more detail below and in Ref.~\cite{Gorshkov2007a}, 
these processes are related by a time reversal-symmetry. This means that the dynamics of either of them fully characterizes the dynamics of the other. For the sake of clarity,  we here introduce both, alongside the relevant notation that will be used in this work. Later, we focus on the retrieval process and resort to the time-reversal symmetry to determine the efficiency and other properties of the storage.

Mathematically, we define $t_1$ and $t_2$ to be the extrema of the time interval over which shape $\phi_s$ is defined, 
i.e., $\phi_s(t) = 0$ when $t \notin [t_1, t_2]$ \footnote{More precisely, $\int_{t_1}^{t_2}{\rm d}t|\phi_s(t)|^2= 1-\varepsilon
$ with $0\le \varepsilon\ll 1$, see also \cite{Giannelli2018}. For practical purposes, in this work we will treat $\varepsilon
$ as it were zero.
}.
Less formally, for storage they are the instants at which the first and last parts of an incoming single-photon pulse reaches the cavity. For retrieval, they indicate the beginning of pulse $\Omega(t)$ and the instant the last part of the excitation leaves the cavity.
As such, $t_2 - t_1$ is the pulse length, and consequently the time required for storing or retrieving the photon into the atoms. 
In this work, we consider both the cases of infinite intervals $t_2 - t_1 \to \infty$, and the experimentally more relevant one where the pulse length $t_2 - t_1$ is finite.

In terms of system operators, we define the efficiencies of the storage and retrieval processes $\eta_s$ and $\eta_r$, respectively, to be given by
%
\begin{subequations}
\label{eq:eta-def-gen}
\begin{align}
\eta_s
&=
\frac{\left\langle \hat{S}^\dagger(t_2)\hat{S}(t_2)\right\rangle}
{\left\langle\hat{\phi}_{s}^\dagger\hat{\phi}_{s}\right\rangle}
,\label{eq:eta_s-def-gen}
\\
\eta_r 
& 
=\frac{\left\langle\hat{\phi}_{r}^\dagger\hat{\phi}_{r}\right\rangle}{\left\langle \hat{S}^\dagger(t_1)\hat{S}(t_1)\right\rangle
}
,\label{eq:eta_r-def-gen}
\end{align}
\end{subequations}
where $\hat{\phi}_{s}$ is the incoming field, given in Eq.\ \eqref{eq:phi_s}, and $\hat{\phi}_{r}$ is the optical outgoing field operator of a desired output mode $\phi_r(t)$, i.e.,
%
\begin{equation}\label{eq:phi_r}
    \hat{\phi}_r = \int_{t_1}^{t_2} 
    \phi_{r}^{*}(t)\hat{\mathcal{E}}_\text{out}(t) dt.
\end{equation}
Similar to $\hat{\phi}_s$, 
for a normalized mode function $\phi_r(t)$, the operator $\hat{\phi}_r$ is a well defined quantum operator with proper commutation relations.
We remark that, while the retrieval efficiency $\eta_r$ is often defined as the total number of retrieved excitations 
(see, e.g., Refs. \cite{Gorshkov2007Universal,Gorshkov2007a,Gorshkov2007b}), we consider the retrieval into a specific mode $\phi_r$, as this will later allow us to relate the storage and retrieval efficiencies. 
In other words, we want to extract efficiently the stored excitations. This ideally implies to retrieve the stored excitations in a specific, target mode.

In typical experimental settings, the average number of thermal photons at optical wavelengths is practically zero. Therefore, we consider the reservoirs associated to $\hat{\mathcal{E}}$ and $\hat{P}$ to be empty, and the noise operators $\hat{\mathcal{E}}_\text{loss}$ and $\hat{F}_{ge}$ in Eqs.~\eqref{eq:eqm} to solely describe vacuum fluctuations. In Ref.~\cite{Gorshkov2007Universal} it was argued that for this situation the performance of a quantum memory can be fully described by its quantum efficiency. Therefore, in this paper we will determine the conditions for which the efficiency is maximal. 

Due to the linearity of the equations of motion Eqs.~\eqref{eq:eqm}, the transformations between the optical fields and the spin-wave mode can be described by the linear mappings $\psi_{s}\left[t,\Omega\left(t\right)\right]$ and $\psi_{r}\left[t,\Omega\left(t\right)\right]$ during storage and retrieval, respectively, such that we have the relations 
\begin{subequations}
\label{eq:map_def}
\begin{align}
\hat{S}(t_2)
&=
\int_{t_1}^{t_2} \psi^*_{s}\left[t,\Omega\left(t\right)\right]\hat{\mathcal{E}}_\text{in}(t)dt
,\label{eq:map_def-storage}\\
\hat{\mathcal{E}}_\text{out}(t) 
&=
\psi_{r}\left[t,\Omega\left(t\right)\right]\hat{S}(t_1)
,\label{eq:map_def-retrieval}\\
 \psi_{r}\left[t,\Omega\left(t\right)\right] 
& = 
\psi_{s}^*\left[t_2-\left(t-t_1\right),\Omega^*\left(t_2-\left(t-t_1\right)\right)\right]
.\label{eq:map_def-time_rev}
\end{align}
\end{subequations}
The first expression describes how an excitation (in the mode  $\phi_{s}$) of the incident field $\hat{\mathcal{E}}_\text{in}$ is stored into the atomic subsystem $\hat{S}$. 
The second equation, on the other hand, stresses that if we extract the excitation from the atomic subsystem, the resulting output field $\hat{\mathcal{E}}_\text{out}$ will be in a certain optical mode characterized by $\psi_{r}$. Notice that, to preserve unitarity, these functions have to fulfill $\int_{t_1}^{t_2}\abs{\psi_s(t)}^2 dt \leq 1$ and  $\int_{t_1}^{t_2}\abs{\psi_r(t)}^2 dt \leq 1$.
Finally, 
the last expression Eq.~\eqref{eq:map_def-time_rev} relates the mappings $\psi_{r}$ and $\psi_{s}$ when the drive used during retrieval is the time-reverse of the drive used for storage (or vice versa; more details below).

We remark that we have for simplicity omitted all vacuum noise operators, since these will vanish once we calculate normal order products, such as the efficiency. Furthermore, we note that the solutions of linear differential equations written in this form are identical regardless of whether they are operator equations or complex number equations. We can  therefore ignore the operator character and simply consider the amplitudes of the various quantum states. This allows us to consider Eqs.~\eqref{eq:in-out} and \eqref{eq:eqm} as standard equations for complex quantities and not operator equations,   
i.e.  replacing  the operators $\hat{S}$, $\hat{P},$ $\hat{\mathcal{E}}_\text{in}$ and $\hat{\mathcal{E}}_\text{out}$ by complex functions $S$,  ${P},$ ${\mathcal{E}}_\text{in}$ and ${\mathcal{E}}_\text{out}$. We will thus omit all hats on operators from now on and simply treat all operators as complex numbers and functions.

Equation~\eqref{eq:map_def-time_rev} follows from the time-reversal symmetry of the Hamiltonian describing the atom-light interaction \cite{Gorshkov2007b}.
This symmetry implies the existence of a one-to-one correspondence between retrieval and time reversed storage. If the first process maps a stored spin-wave excitation into a mode defined by $\psi_{r}$, the time reversed storage process will map the time reversed mode into the spin wave $S$. There is thus only a single relevant mode function and for ease of notation we omit the subscripts in the following and define $\psi\left[t,\Omega\left(t\right)\right]=\psi_{r}\left[t,\Omega\left(t\right)\right]$ and $\phi(t) = \phi_r(t)$ such that
\begin{subequations}\label{eq:mapping_psi}
\begin{align}
&S(t_2) = 
\int_{t_1}^{t_2} \psi\left[t_2-\left(t-t_1\right),\Omega^*\left(t_2-\left(t-t_1\right)\right)\right]\mathcal{E}_\text{in}(t) dt
,\label{eq:S_mapping}
\\
&\mathcal{E}_\text{out}(t) = 
\psi\left[t,\Omega\left(t\right)\right]S(t_1)
.\label{eq:E_out_psi_S}
\end{align}
\end{subequations}
%

Considering now the retrieval of a normalised spin wave $\abs{S(t_1)}^2=1$ and the storage of a normalised mode 
\begin{equation}
\mathcal{E}_\text{in}(t) = \phi_{s}(t) = \phi^*(t_2-(t-t_1))
\label{eq:storagemode}
\end{equation}
with $\int_{t_1}^{t_2}\abs{\phi_s(t)}^2dt=1$, we can insert Eqs.~\eqref{eq:mapping_psi} into Eqs.~\eqref{eq:eta-def-gen} to rewrite the storage and retrieval efficienies as 
\begin{subequations}
\label{eq:eta}
\begin{align}
\eta_s 
& =
\absa{\int_{t_1}^{t_2} \psi\left(t_2-\left(t-t_1\right)\right)\phi^*\left(t_2-\left(t-t_1\right)\right)dt}^2
, \label{eq:eta_s} \\
\eta_r & = 
\absa{ \int_{t_1}^{t_2} \phi^*(t)\psi(t)dt}^2
,\label{eq:eta_r}
\end{align}
\end{subequations}
%
which demonstrates that they are equal to each other.

From Eq.~\eqref{eq:eta_s}, 
it follows that to find the optimal $\eta_s$ we must  find a mapping $\psi$ that maximizes its overlap with $\phi$. According to the time reversal argument, this means that the optimal storage is the time reverse of the optimal retrieval into the time reversed mode 
and these two processes have the same efficiency \cite{Gorshkov2007b}. From their equivalence, we can choose to optimize either one of them. The retrieval process is technically simpler since $\mathcal{E}_\text{in}=0$ simplifies the calculations. We will therefore perform all the detailed derivations for the retrieval process. The incoming mode function $\phi_s(t) = \phi^*(t_2-(t-t_1))$ is then only necessary for defining the desired output shape through Eq.~\eqref{eq:storagemode}.
For ease of notation, we  drop subscripts in the following when referring to, e.g., the efficiencies.

\subsection{Remarks on the ``atom-limited'' and ``cavity-limited'' regimes}
While the memory efficiencies can be numerically evaluated and optimized 
\cite{Giannelli2018}, we look for  analytical solutions that can be used for designing pulses $\Omega(t)$ that improve the quality of the quantum memory. For this purpose, 
it is desirable to separate the analysis into the ``cavity-limited" 
$\Gamma \pt \gg 1$
and ``atom-limited" 
$\kappa \pt \gg 1$
regimes, where $\pt$ is a characteristic timescale
\footnote{Note that, generally, there may be several timescales characterizing a pulse to be stored. For simplicity and clarity, in this work we restrict ourselves to a single time. In practical settings, one should identify and employ the one(s) that are relevant for storage and retrieval. However, as better explained in Secs.~\ref{sec:adiabatic_elimination_bad_cavity} and \ref{sec:adiabatic_elimination_good_cavity}, our approaches always yield an experimentally viable strategy that can be implemented in the setup under consideration.}
for shape $\phi$ (see, e.g., Table~\ref{table:inf_pulses}) and 
\begin{equation}
 \Gamma = \frac{g^2 N}{\kappa_{\rm in}}  
\end{equation}
is the effective radiative linewidth due to coupling with the cavity.
The ``atom-limited'' case will be considered in Sec.~\ref{sec:Bad_cavity_limit}, while we defer the analysis of the ``cavity-limited'' case to Sec.~\ref{sec:Good_cavity_limit}. 

The ``cavity-limited'' regime is defined by the inequality $\Gamma \pt \gg 1$. 
This means that the response time of the atomic medium is sufficiently short
to perfectly store the excitations transferred to the cavity field $\mathcal{E}$. In this regime the storage efficiency is 
limited by the rate at which the cavity can accommodate the incoming pulse $\phi$. Conversely, if $\kappa \pt \gg 1$ the cavity response time is fast enough to capture incoming excitations and the storage efficiency is limited by the characteristic timescale of the atomic ensemble.

These two complementary regimes are somewhat related to the so-called
``bad cavity" $\kappa/(g\sqrt{N}) \gg 1$ \cite{Sorensen2002Entangling} and ``good cavity" $\kappa/(g\sqrt{N}) \ll 1$ \cite{Tolazzi2021} limits. In our case, we believe that the ``atom-limited'' $\kappa \pt \gg 1$ and the complementary ``cavity-limited'' $\Gamma \pt \gg 1$ regimes better capture the memory response. In fact, we will show that they allow us to better understand the analytical results and also to shed light on the analytical procedure we implement in
Secs.~\ref{sec:adiabatic_elimination_bad_cavity} and \ref{sec:adiabatic_elimination_good_cavity} below. 
Indeed, in the context of atomic memories, what discriminates between the ``atom'' and ``cavity-limited'' regimes is not (directly) the ratio between the rates at which light enters the cavity and excitations are transferred between the cavity and atomic modes, respectively. Instead, what really matters is (1) whether the cavity is fast enough to absorb the incoming pulse, thereby suppressing reflection 
and (2) whether the atoms can transfer the photon to be stored sufficiently fast to match the desired intensity of its pulse shape. These two conditions clearly depend on the parameters of the system, but also on the properties of $\phi$, in particular its characteristic timescale $\pt$. In fact, for any given memory, it is always possible to find exotic input pulses that cannot be efficiently stored. Therefore, for the ``atom-limited'' regime, we study the scenario in which the cavity (1) is not the limiting factor, and all detrimental contributions to the efficiency arise from the atomic medium. Vice versa, in the ``cavity-limited" regime the atomic medium (2) is not the limiting factor, and the detrimental contributions are due to the cavity.

An interesting consequence from our definitions of ``atom'' and ``cavity-limited'' regimes is that a memory can operate in both, as long as the identifying limits are simultaneously fulfilled. 
As discussed in Sec.~\ref{sec:Bad_cavity_limit} below, this corresponds to the adiabatic regime already investigated in Ref.~\cite{Gorshkov2007a}. In this scenario, either of the strategies developed in this work for the ``cavity'' and ``atom-limited'' regimes work and yield the same efficiency.

\section{``atom-limited" regime}
\label{sec:Bad_cavity_limit}
In the ``atom-limited" case $\kappa \pt \gg 1$, the coupled atom-cavity dynamics is dominated by the atoms, and for retrieval (such that $\mathcal{E}_{\text{in}}=0$) Eqs.~\eqref{eq:eqm} can be simplified by setting $\dot{\mathcal{E}} \simeq 0$ and adiabatically eliminating \cite{Lugiato1984Adiabatic} the cavity mode $\mathcal{E}$. 
The validity of this approximation 
will be investigated in detail in Sec.~\ref{sec:adiabatic_elimination_bad_cavity}. 
%
The cavity field thus becomes
\begin{equation}\label{eq:bcl_efield}
    \mathcal{E} = \frac{i\sqrt{N}g}{\kappa} P,
\end{equation}
which can be substituted back into Eqs.~\eqref{eq:eqm} to obtain
\begin{subequations}
\label{eq:gen_bc}
\begin{align}
\mathcal{E}_\text{out}&=\frac{\kappa_\text{in}-\kappa_\text{loss}}{\kappa}\mathcal{E}_\text{in}+i\sqrt{2\tilde{\Gamma}}\sqrt{\frac{C}{1+C}}\sqrt{\frac{\kappa_\text{in}}{\kappa}}P,\label{eq:gen_bc-E_out}\\
\dot{P}&=-\left[\tilde{\Gamma}+i\Delta\right]P+i\Omega S+i\sqrt{2\tilde{\Gamma}}\sqrt{\frac{C}{1+C}}\sqrt{\frac{\kappa_\text{in}}{\kappa}}\mathcal{E}_\text{in},\label{eq:gen_bc-P_dot}\\
\dot{S}&=i\Omega^*P.\label{eq:gen_bc-S_dot}
\end{align}
\end{subequations}
In these equations, we have introduced the total atomic decay rate $\tilde{\Gamma} = \gamma + \kappa_{\rm in} \Gamma / \kappa = \gamma\left(1+C\right)$, consisting of the decay rate into free space $\gamma$ and through the cavity $g^2N/\kappa$. The ratio between these decay rates is the cooperativity $C=g^2N/(\kappa\gamma)$ \cite{Gorshkov2007a}. In the lossless case, where $\kappa_\text{loss}=\gamma=0$, $\kappa = \kappa_{\rm in}$ and $\tilde{\Gamma} = \Gamma = g^2N/\kappa$, we can rewrite Eqs.~\eqref{eq:gen_bc} as
\begin{subequations}
\label{eq:lossles_bc}
\begin{align}
\mathcal{E}_\text{out}^0 &= \mathcal{E}_\text{in}+i\sqrt{2\Gamma}P^0, \label{eq:lossles_bc-E_out}\\
\dot{P}^0 &= -\left[\Gamma+i\Delta\right]P^0+i\Omega S^0+i\sqrt{2\Gamma}\mathcal{E}_\text{in}, \label{eq:lossles_bc-P_dot}\\
\dot{S}^0 &= i\Omega^*P^0, \label{eq:lossles_bc-S_dot}
\end{align}
\end{subequations}
where subscript ``0" refers to the lossless case.
Physically, the last set of equations describes a three-level system that can either decay into or be driven from a waveguide, both with a coupling rate $\Gamma$. In this case, 
excitations are never lost due to dissipation, $\gamma \rightarrow 0$ and $\kappa_\text{loss}\rightarrow 0$, making the cooperativity infinite, $C\rightarrow \infty$, and $\kappa \rightarrow \kappa_\text{in}$. In Eqs.~\eqref{eq:gen_bc}, on the other hand, nonzero values of $\gamma$ and $\kappa_\text{loss}$ yield finite cooperativity $C$ and consequently loss of the photon (excitation) during the storage (retrieval) process.

Remarkably, it is possible to express the efficiencies $\eta_s$ and $\eta_r$ achieved in the lossy case in terms of the corresponding quantities $\eta_s^0$ and $\eta_r^0$ obtained for the lossless case. During retrieval, $\mathcal{E}_\text{in} = 0$ and the two systems of equations~\eqref{eq:gen_bc} and ~\eqref{eq:lossles_bc} become equivalent once $\mathcal{E}_\text{out}$ is mapped onto $\mathcal{E}_\text{out} \rightarrow \sqrt{C/(1+C)}\sqrt{\kappa_\text{in}/\kappa}\:\mathcal{E}^0_\text{out}$ \cite{Giannelli2018}. Similarly, for  the storage process, 
we can compare the terms that are driven by  the source term $\mathcal{E}_\text{in}$ in Eqs.~\eqref{eq:gen_bc} and Eqs.~\eqref{eq:lossles_bc} and  see that they differ by the same factor as found in the retrieval process. From the definition of the efficiencies $\eta_s$ and $\eta_r$ in Eqs.~\eqref{eq:eta-def-gen}, we thus conclude that
%
\begin{subequations}
\label{eq:efficiencies_loss_lossless}
\begin{align}
    \eta_s&=\frac{C}{1+C}\frac{\kappa_\text{in}}{\kappa}\eta_s^0,\\
    \eta_r&=\frac{C}{1+C}\frac{\kappa_\text{in}}{\kappa}\eta_r^0,
\end{align}
\end{subequations}
meaning that, in the ``atom-limited" case, the lossless case suffices for deriving a complete description of the effect of the pulse shape; see also \cite{Giannelli2018}. The losses will just enter as a constant factor in front of the efficiencies, provided that the comparison is performed with the same total decay rates $\Gamma$ and $\kappa$ in the two cases.  
In the following derivations for the ``atom-limited" case, we thus restrict ourselves to the lossless case, omitting the superscript ``0" for brevity. 

The (lossless) equations of motion~\eqref{eq:lossles_bc} describe how an excitation stored into $P$ decays into the outgoing mode $\mathcal{E}_\text{out}$ at a rate $\Gamma$, and how the input field $\mathcal{E}_\text{in}$ is mapped onto $P$, again controlled by the rate $\Gamma$. 
%
%
The effective decay rate $\Gamma$ therefore 
controls the bandwidth of the memory, i.e. the frequency range of the input photon that can be stored without losses  in the system. 

For a sufficiently long pulse length $\Gamma \pt\gg 1$, i.e. when we also meet the requirement of the ``cavity-limited'' regime, we can also adiabatically eliminate $P$ to simplify the treatment even further. In this limit -- known as the adiabatic regime -- it was shown in Ref.~\cite{Gorshkov2007a} that, by appropriately shaping $\Omega(t)$, it is possible to obtain unit storage and retrieval efficiencies $\eta_s=\eta_r=1$ in the lossless case for arbitrary input $\phi_s$ and output $\phi_r$ mode shapes
\footnote{
To be precise, to fully meet the requirements of the adiabatic regime, we also need the incoming pulse shape $\phi$ to be of infinite duration. In Sec.~\ref{sec:finite_pulses} we demonstrate that any finite pulse cannot be perfectly stored due to its tails, and identify their characteristics that are most detrimental to $\eta$ (such that they can be avoided or suppressed).
}. This is not true in the nonadiabatic regime $\Gamma \pt \lesssim 1$, where detrimental effects become unavoidable even for lossless  memories. In the following section, we develop a procedure that allows us to maximize the efficiency even in this nonadiabatic regime for a lossless memory. More precisely, we determine a shape for $\Omega(t)$, which is arguably optimal.

\subsection{Optimal memories in the nonadiabatic regime}
\label{sec:optimal_memories}

Several studies have devised protocols to optimize quantum memories \cite{Giannelli2018,Bensky2012Optimizing,saglamyurek2018coherent,Asenjo2017Exponential}. By either working in the adiabatic regime ($\Gamma \pt\gg 1$ \textit{and} $\kappa \pt \gg 1$) \cite{Gorshkov2007a} or requiring impedance matching such that the cavity has no reflection during storage \cite{Dilley2012Single}, it is possible to find strategies with unit efficiencies.
These studies, however, do not yield the best achievable outcome for short pulses, when the adiabatic approximation breaks down. In this regime, so far, optimization is achieved by means of numerical approaches \cite{Giannelli2018,Shinbrough2021,Novikova2007}. In this section, we develop an analytical method for efficiently operating a quantum memory inside \textit{and} outside the adiabatic regime. 

\subsubsection{Optimization of $\psi$}
\label{sec:optimal_memories_psi_optimization}
%
%
\begin{figure*}
	\centering
	\includegraphics[width=\textwidth]{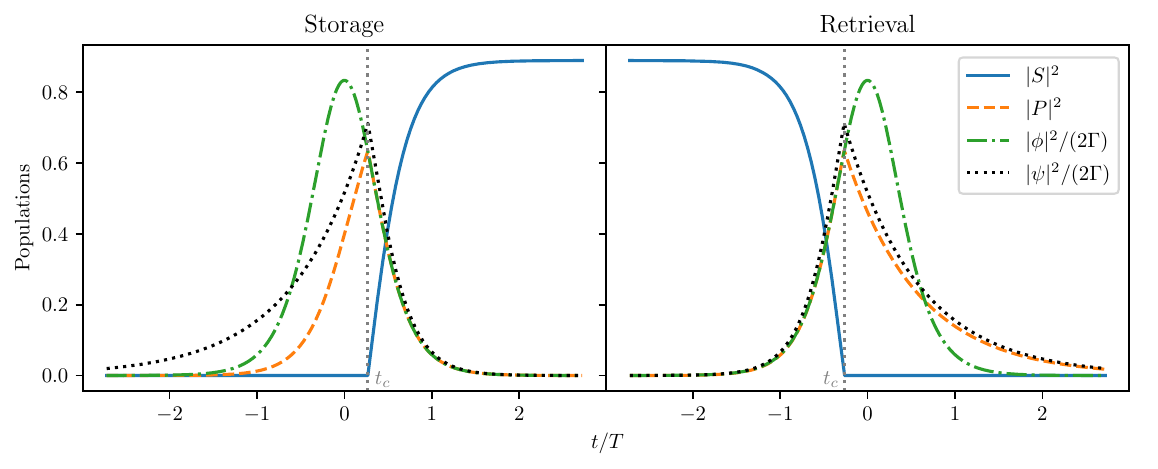}
	\caption{
Time evolution of a storage followed by retrieval processes, for a sech pulse shape $\phi(t)$ (green dot-dashed lines; see Table~\ref{table:inf_pulses} in Sec.~\ref{sec:Infinite_Pulses}) and optimized via our protocol. The mapping $\psi(t)$ (black dotted lines) is obtained via Eq.~\eqref{eq:Psi}, resulting in the operators $S(t)$ and $P(t)$ (solid blue and dashed orange lines, respectively). The critical times $t_c$ are found following the procedure in the main text, and are highlighted by the vertical gray dotted lines. The fields $\phi$ and $\psi$ have been rescaled to have the same units as $P$ and $S$. Notice that since we consider the combined storage and retrieval process to illustrate the time reversal property of the dynamics after (before) $t_c$ for storage (retrieval), the value of $|S|^2$ at the beginning of the retrieval process is the same as at the end of the storage process. We have chosen the parameters $\Gamma \pt=0.6$ and $\left|t_2-t_1\right|=\sqrt{3}\pi \pt$.
 }
	\label{fig:evolution}
\end{figure*}
%

For ideal systems in the adiabatic regime ($\Gamma \pt \gg 1$), we argue below that we can always chose $\psi(t)=\phi(t)$ such that 
$\eta = 1$.
There are, however, physical constraints on the choice of $\psi$ that prevent reaching unit efficiency in the nonadiabatic regime. 
%
%
%
According to the relation $\mathcal{E}_\text{out}=i\sqrt{2\Gamma}P$ from Eq.~\eqref{eq:lossles_bc-E_out} with $\mathcal{E}_{{\rm in}}=0$, one can control the outgoing photon flux during the retrieval by adjusting the population in the excited state $\ket{e}$. The amount of population that can be stored in the atomic medium is, however, bounded and that puts a limitation on $P$. Once the excitation is transferred to the upper level $\ket{e}$, the outgoing flux is limited by the probability of a spontaneous decay, which is the probability of occupation of the excited state multiplied by the decay rate $\Gamma$. 
For short pulses, $\Gamma \pt\lesssim 1$, unit retrieval efficiency to a certain mode cannot be achieved. In fact, the desired field mode may require a larger flux than what can be obtained from the memory because the excitation is retrieved at the finite decay rate $\Gamma$. This observation is the key to the optimization procedure we implement below, where this effect is found to be the only limiting factor for the efficiency.

Starting from these qualitative observations, we will now determine a mathematical constraint
on $\psi$. In the ``atom-limited" case the field $\mathcal{E}$ satisfies Eq.~\eqref{eq:bcl_efield}, and we can therefore neglect its population (since $\kappa \pt \gg 1$ implies that $P/\kappa$ is negligible).
Conservation of the total number of excitations during retrieval then implies that
\begin{equation}\label{eq:conservation}
\abs{S(t)}^2+\abs{P(t)}^2+\int_{t_1}^{t}\abs{\mathcal{E}_\text{out}(t')}^2dt'=\abs{S(t_1)}^2
\end{equation}
for all instants of time $t\ge t_1$. Here, the left hand side describes all excitations that are either retained in the atomic ensemble or have left through the outgoing mode $\mathcal{E}_\text{out}$. 
Since we assume that $\kappa_{\rm loss}=0$, the left hand side of Eq.~\eqref{eq:conservation} must be equal to the number of excitations $\abs{S(t_1)}^2$ at the beginning of the retrieval process. We remark that when considering the retrieval process alone, $\abs{S(t_1)}^2$ is set to one. In the following equations, we keep $\abs{S(t_1)}^2$ as an unspecified (positive) parameter thereby including the description of the composite storage plus retrieval process. 

$P$ and $\mathcal{E}_\text{out}$ are related to $\psi$ by Eq.~\eqref{eq:lossles_bc-E_out} and Eq.~\eqref{eq:E_out_psi_S} such that
\begin{equation}\label{eq:S_psi_inequality}
\frac{\abs{S(t)}^2}{\abs{S(t_1)}^2}+\frac{1}{2\Gamma}\abs{\psi(t)}^2+\int_{t_1}^t\abs{\psi(t')}^2dt'=1
\end{equation}
for all $t$. By using the  constraint $\abs{S(t)}^2\geq 0$, 
we can thus obtain an inequality 
%
\begin{equation}\label{eq:psi_inequality}
\abs{\psi(t)}^2\leq 2\Gamma\left(1-\int_{t_1}^t\abs{\psi(t')}^2dt'\right),
\end{equation}
%
setting a limit on the maximal retrieval rate. 
The physical meaning of 
this last equation is that the flux out of the system is limited by the effective decay rate $\Gamma$ times the population left in the system.
Since the desired outgoing field $\phi$ may have an intensity that exceeds the bound in Eq.~\eqref{eq:psi_inequality}, we cannot always match the shape of the pulse $\phi$, and in turn this limits the efficiency $\eta$ that can be achieved. 
Our goal is thus to optimize $\psi$ such that the constraint in Eq.~\eqref{eq:psi_inequality} is satisfied while maximizing the scalar product in Eq.~\eqref{eq:eta_r}.

The optimization strategy we propose is the following. During the retrieval process, we fix the shape $\psi(t)$ such that, for $t \in [t_1,t_c]$, it is equal to the desired shape $\phi(t)$ up to a constant $c$. Here, we have introduced the time $t_c \leq t_2$, which we will denote by ``critical time". At this critical time $t_c$ all excitations have been transferred out of the state $\ket{s}$, i.e., $\abs{S(t_c)}^2=0$. Thus, at $t_c$ the inequality in Eq.~\eqref{eq:psi_inequality} saturates, and since we cannot further increase the output rate, the dynamics can no longer follow the desired shape $\phi (t)$. 
For larger times, we want the system to be as close as possible to the (unattainable) desired output intensity. This means that we want the maximal flux out of the system and we obtain this by leaving all excitations in the excited level $\ket{e}$. Therefore, after the critical time $t_c$, $\psi(t)$ shall be just given by  the exponential decay of the excitations out of $P$. 

By putting these observations together, we can express our ansatz for the mode function $\psi(t)$ as
\begin{equation}\label{eq:Psi}
\psi(t)= \begin{cases}
c\:\phi (t), &t \leq t_c\,,\\
c\:\phi (t_c)e^{-\Gamma\left(t-t_c\right)}, &t>t_c\,.
\end{cases}
\end{equation}
We remark that, for simplicity, we ignore the possibility that the desired output state may have a phase evolution after $t_c$ 
\footnote{
To implement such a phase evolution, one can employ a far off-resonant drive $\Omega(t)$ that induces an AC-Stark shift of $\ket{e}$, which adds a phase to $\psi(t)$
}. 

The advantage of having the parameter $c$ in Eq.~\eqref{eq:Psi} can be understood from the saturation of the inequality in Eq.~\eqref{eq:psi_inequality}. This limitation on the mapping $\psi$ implies that the total outgoing photon flux will be reduced after $t_c$ compared to the desired pulse shape (see Fig.~\ref{fig:evolution}). We compensate for this by including the constant $c\geq 1$, which allows increasing the intensity before $t_c$ thus achieving a larger overlap with the desired shape $\phi$ for $t\in[t_1,t_c]$. We remark that the parameter $c$ is upper bounded by the maximal retrieval rate from the system, which depends on the amount of excitations left at any given time. This limitation practically plays a role only when the critical and initial times are the same $t_c = t_1$, as for the exponentially increasing pulse in Sec.~\ref{sec:Infinite_Pulses}.

Our strategy is illustrated in Fig.~\ref{fig:evolution}, where we show the dynamics of the storage and retrieval process, optimized with our method for a hyperbolic secant (sech) pulse as input and desired output.
In the figure, we show the evolution of $|S|^2$, $|P|^2$ (full blue and dashed orange lines, respectively) and $|\psi|^2$ (black, dotted line), and compare them with the incoming pulse $|\phi|^2$ (green dot-dashed line). The critical time $t_c$ is highlighted by the vertical dashed line.

As a sanity check for our physically motivated protocol, we 
demonstrate that the shape $\psi$ in Eq.~\eqref{eq:psi_inequality} has a number of salient features: First, 
 it yields unit storage and retrieval efficiencies in the adiabatic regime. 
Indeed, when $\Gamma \pt \gg 1$, the intensity is small 
compared to the timescale $\Gamma^{-1}$ at which the system can store/retrieve excitations, and we can always have $\psi(t) = \phi (t)$ in Eq.~\eqref{eq:Psi} while satisfying the constraint in Eq.~\eqref{eq:psi_inequality} (since $|\psi|^2\sim 1/\pt$). 
Secondly, for  pulses close to the adiabatic threshold $\Gamma \pt \approx 1$, near-optimal efficiencies are obtained  by shaping the retrieval $\psi$ such that it matches that desired shape 
$\phi (t)$ as closely as possible. 
We achieve this by ensuring that, up to a constant $c$ [see Eq.~\eqref{eq:Psi}], $\psi(t)$ equals the desired shape $\phi (t)$ for as long as possible, i.e., until the critical time $t_c$ at which the system exhausts the excitations stored in $\ket{s}$. After $t_c$, the best strategy is to have the highest possible output, that is, follow the exponential decay out of $P$.


In App.~\ref{app:c_opt} we prove that  $\eta$ 
is optimized with respect to the constant $c$ when $c=1/\sqrt{\eta}$. 
%
Incidentally, this value for $c$ means that the best strategy for the combined process of storage followed by retrieval is the one where the outgoing pulse shape is identical to the time reversed incoming shape before $t_c$,
i.e., $\mathcal{E}_\text{out}(t)=\phi (t)$ for $t<t_c$. 
Within this limit, this further 
ensures that both $|S|^2$ and $|P|^2$ are symmetric (see  Fig.~\ref{fig:evolution}). Outside this regime, on the other hand, excitations are reflected from  the system during storage
and the output field does not match the input during retrieval, making the memory imperfect. Remarkably, this means that the condition under which the efficiencies $\eta_s$ and $\eta_r$ are maximized, is that the optical fields $\mathcal{E}$ [equivalently $P$, see Eq.~\eqref{eq:bcl_efield} and Fig.~\ref{fig:evolution}] inside the cavity during the combined storage and retrieval processes, are time reversals of each other in an interval set by $t_c$ (but not outside). At the moment we do not have a simple physical explanation for why this is the case. We emphasize that this symmetry is not the same as the general time reversal symmetry of the equations that we use for the optimisation.  

In Refs.~\cite{Vasilev2010,Utsugi2022,Tissot2024}, a similar optimization method as the one we use here was considered. In these works it is also noted that ideal performance can be achieved if there is no violation of the inequality in Eq.~\eqref{eq:psi_inequality}. When ideal performance cannot be achieved, however, their strategy consists in lowering the value of $c$ to less than unity. 
Since the optimal performance is achieved for $c=1/\sqrt{\eta}\geq 1$ (see App.~\ref{app:c_opt}), this strategy necessarily leads to a worse performance than what is achieved here. A more detailed comparison is presented in App.~\ref{app:comparison}.


%
\subsubsection{Comparison to numerical optimizations}
\label{sec:optimal_memories_numerics}

%
%
\begin{figure}[]
	\centering
        \subfigure{\includegraphics[width=\columnwidth]{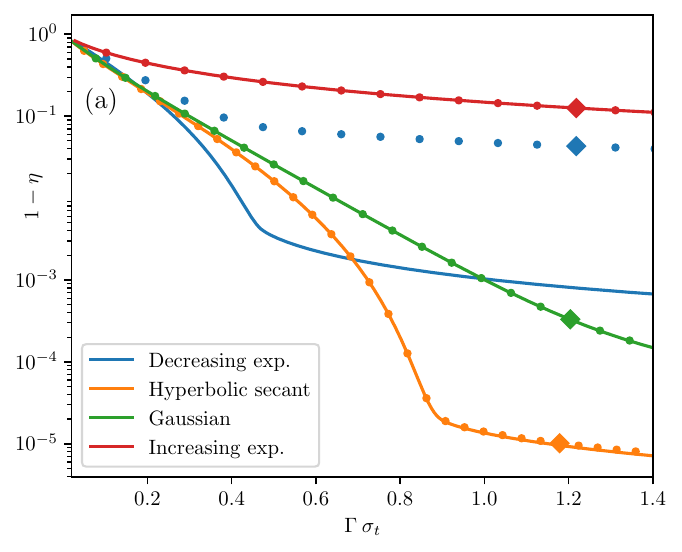}}
        \subfigure{\includegraphics[width=\columnwidth]{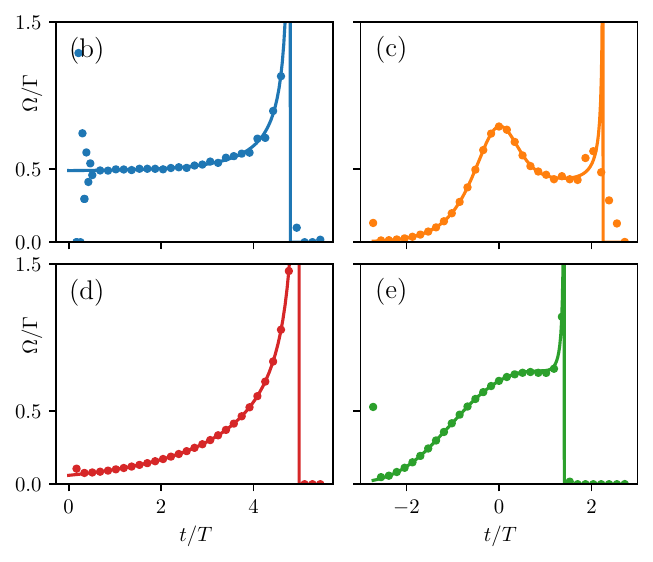}
        \label{fig:Optimization_Omega}}%
	\caption{
 Comparison between our protocol (solid lines) and an OCT numerical optimization (dots) for the pulses in Table~\ref{table:inf_pulses} (as indicated in the legend; the Lorentzian shape is investigated in Fig.~\ref{fig:two_tc}). In (a), we present the inefficiencies for different values of $\Gamma \ct $, where $\ct $ is the time variance defined in Eq.~\eqref{eq:coh_time}. Our method employs the ansatz in Eq.~\eqref{eq:Psi} with $c$ determined by a fixed-point iteration, while the numerical OCT method optimizes directly over $\Omega(t)$ with Eqs.~\eqref{eq:lossles_bc} as dynamical equations. In (b)-(e), we present the shapes of $\Omega(t)$ corresponding to the values marked with a diamond shape in (a) at $\Gamma \ct  \approx 1.2$. For our protocol, $\Omega(t)$ is found from $\psi(t)$ following the strategy presented in Sec.~\ref{sec:omega_determination}. In both panels, we set $\Delta = 0$, and truncate and renormalize $\phi$ such that $|t_2-t_1| = \sqrt{3} \pi \pt$ and $\int_{t_1}^{t_2}|\phi|^2(t) dt = 1$. We remark that, due to the finite interval, values of $\ct $ are not the ones in Table~\ref{table:inf_pulses}, but are calculated directly from Eq.~\eqref{eq:coh_time}. Furthermore, we highlight that for the decreasing exponential in (b) our protocol prescribes an additional delta-function in $\Omega$ at $t=0$ due to the abrupt rise in the shape  $\phi$. As a consequence of finite time intervals, a similar contribution occurs for the other shapes when optimized with truncated pulses.
}
\label{fig:Optimization}
\end{figure}

Our method  determines a physically motivated shape $\psi$ through an optimization over the single parameter $c$. 
With our procedure, $c$ is determined via the relation $c=1/\sqrt{\eta}$, satisfied when optimality is achieved (see Appendix~\ref{app:c_opt}). We can therefore employ a fixed-point iteration to determine its value numerically \cite{Hoffman2001}. We calculate the efficiency $\eta$ with one value of $c$ and use the resulting $\eta$ to determine the value of $c$ in the next iteration. We then repeat the process until convergence.

An important question to address is how well this strategy performs. Before going into the details of the method, we start by comparing the performance of our protocol  
with a numerical optimization obtained with optimal control theory (OCT) \cite{KochEQT2022quantum,WilhelmAQ2020introduction,giannelli2022tutorial,Goerz2019Krotov}, which consists in numerically maximizing the efficiency $\eta$ as a function of $\Omega(t)$, satisfying the dynamical equations~\eqref{eq:lossles_bc}. The control pulse $\Omega(t)$ is the cubic interpolation between the points associated with the $n=129$ time instants into which $[t_1,t_2]$ is split. These points are then optimized with the limited-memory Broyden-Fletcher-Goldfab-Shanno algorithm \cite{giannelli2022tutorial,byrd1995LBFGSB}.
%
For completeness, we remark that the numerical calculations (both for our protocol and the OCT method) are designed for the storage process. Indeed, while the analytical results are better understood for retrieval, it is slightly more convenient to perform the simulations for storage.

In Fig.~\ref{fig:Optimization}(a), we consider the hyperbolic secant (sech), Gaussian, and rising and decaying exponential pulses $\phi$ (see Table~\ref{table:inf_pulses} in Sec.~\ref{sec:Infinite_Pulses}) and show the inefficiencies $1-\eta$ 
when varying $\Gamma \ct $. As can be seen, the efficiencies obtained by our method (solid lines)
are always comparable or better than the numerical optimization (dots).
In the figure it is clear that some of the pulse shapes have different behaviors in different limits, resulting in a kink in the curves of the efficiency. 
These kinks are explained in Secs.~\ref{sec:Infinite_Pulses} and \ref{sec:finite_pulses} below, where asymptotic limits for the memory's inefficiencies associated with different input shapes $\phi$ and different truncations of the time interval $t_2-t_1$ are derived.

We emphasize that despite the physical motivations behind our method and the strong numerical evidence, we cannot prove that our solution is the true optimum. However, the optimization of $\psi$ subject to the constraint~\eqref{eq:psi_inequality} always leads to a physically acceptable solution, no matter the input shape $\phi$, and we never found instances in which the full numerical optimization outperformed our strategy.   

There are, however, examples where we need to extend the simple strategy outlined so far. 
For pulses with either more than one peak or tails that decrease slower than an exponential, inequality~\eqref{eq:psi_inequality} can be fulfilled again at a later stage of the retrieval. In these cases, we can increase the efficiency $\eta$ by changing the mapping $\psi$ in Eq.~\eqref{eq:Psi} to contain more critical times.
We analyze this for the particular case of a Lorentzian shaped $\phi$ in Sec.~\ref{sec:Infinite_Pulses} below, where we show that the overlap with the slowly decaying tails of $\phi$ can be increased by using more than one $t_c$. 
\subsubsection{Determination of $\Omega$}
\label{sec:omega_determination}
Above, we have only considered shape $\psi$ in the optimization and ignored the driving field $\Omega$. To apply the procedure in experiments, it is, however, essential to know which driving to apply. Importantly, albeit our results refer to the lossless case (which is never experimentally accurate), they can be promptly mapped to the lossy one via the correspondence outlined above Eqs.~\eqref{eq:efficiencies_loss_lossless}. 
Here, we explain how one can always determine $\Omega$ from mapping $\psi$ given by our method. As the derivation of the phase of $\Omega$ is lengthy, for brevity we here only provide the result for $|\Omega|$, and redirect the reader to Appendix~\ref{app:Phase_S} for more details.


From Eqs.~\eqref{eq:lossles_bc} we can isolate $\Omega$, which can be written in terms of $\psi$ only:
\begin{equation}
\label{eq:Omega_st1}
    \Omega(t)=
    \begin{cases}
    -\frac{
    \dot{\psi}(t)+\left(\Gamma+i\Delta\right)\psi(t)
    }{
    \sqrt{2\Gamma-\abs{\psi(t)}^2-2\Gamma\int_{t_1}^t\abs{\psi(t')}^2dt'}
    }
    e^{-i\theta(t)}, & t < t_c,\\
    0, & t \geq t_c,
\end{cases}
\end{equation}
with $\mathcal{E}_\text{in}=0$ since we are considering the retrieval process.
Here, we used $P=\psi S(t_1)/i\sqrt{2\Gamma}$ and determined $|S|$ from Eq.~\eqref{eq:S_psi_inequality}. The phase of $S$ corresponding to $\theta$ in $S=\abs{S}e^{i\theta}$ is derived in Appendix~\ref{app:Phase_S}. It follows from Eq.~\eqref{eq:Omega_st1} that, for any 
$\psi$, we can find the desired 
$\Omega$ for $t<t_c$. At the critical time $t_c$, the atomic level $\ket{s}$ is depleted and we have $|S(t_c)| = 0$, implying that $\Omega(t_c) \overset{+}{\rightarrow} \infty$. After the critical time $t_c$ we want all excitations to remain in the excited state (see Eq.~\eqref{eq:Psi} and the following discussion), and thus set 
$ \Omega(t)=0$ for $t \geq t_c$.

In Fig.~\ref{fig:Optimization}(b-e), we compare the driving obtained with this method to that obtained with OCT. From the figure, it is apparent that the driving shapes are essentially identical. Deviations mainly appear due to abrupt changes in $\phi$ at the onset of the pulse. Here, Eq.~\eqref{eq:Omega_st1} [recall that $\psi \propto \phi$ for $t\leq t_c$; see Eq.~\eqref{eq:Psi}] predicts a delta function in $\Omega$
that is challenging to capture numerically. This leads to a worse performance of OCT for the decreasing exponential, which has a very abrupt change at the onset (see Appendix~\ref{app:decr_exp} for further investigations of this). Note that, since the investigated pulses are truncated on a finite interval, all pulses have  an abrupt onset, but it is much smaller for the other pulse shapes. These delta-function contributions correspond to a rapid pulse exciting the atom(s) to the excited state with a finite probability. Also, deviations appear near the critical time $t_c$, where $\Omega$ diverges due to the denominator in Eq.~\eqref{eq:Omega_st1} -- again hard to capture numerically.

Another feature of $\Omega$ in Eq.~\eqref{eq:Omega_st1} is that, in the adiabatic regime $\Gamma \pt \gg 1$, it reduces to the previously found optimal driving shape in Ref.~\cite{Gorshkov2007a}. Indeed, the limit $\Gamma \pt\gg 1$ allows us to determine $P$ by setting $\dot{P}\approx 0$ in Eq.~\eqref{eq:lossles_bc-P_dot}, and to find $S$
from Eq.~\eqref{eq:S_psi_inequality}. Therefore, in the adiabatic regime the driving 
$\Omega(t)$ becomes (including the phase $\theta$ -- see Appendix~\ref{app:Phase_S})
%
\begin{equation}
\label{eq:Omega_atom_limited_full}
\begin{split}
    \Omega(t)=&-\frac{\Gamma +i\Delta}{\sqrt{2\Gamma}}\frac{\phi(t)}{\sqrt{\int_t^{t_2}\abs{ \phi(t')}^2}dt'} \\
    & \times \exp\left\lbrace -\frac{i\Delta}{\Gamma^2+\Delta^2}\int_{t_1}^{t}\abs{\Omega(t')}^2dt'\right\rbrace.
    \end{split}
\end{equation}
This result coincides with the result obtained in Refs.~\cite{Gorshkov2007a,Giannelli2018} in the adiabatic regime.
\subsection{Infinite time intervals}
\label{sec:Infinite_Pulses}
\begin{table*}
$\begin{array}{l!{\vrule width 2pt}l|l|l|l|l|l}
  \textbf{Retrieval shape}
  & \phi(t)
  & \ct 
  & [t_1,t_2] 
  & T_{\rm a}\:[\Gamma^{-1}] 
  & 1-\eta\:\text{for}\: \pt\rightarrow T_{\rm a} 
  & 1-\eta\:\text{for}\:\pt\rightarrow \infty
  \\ \hline \hline 
 \textit{Decreasing exp.} 
 & \frac{1}{\sqrt{\pt}}\exp\left(-\frac{t}{2\pt}\right)
 & \pt
 & \left[0,\infty\right[ 
 & \frac{1}{2} 
 & \Gamma^2\left(T_{\rm a}-\pt\right)^2
 & 0 
 \\ \hline
 \textit{Hyperbolic secant} 
 &\frac{1}{\sqrt{\pt}}\sech{\frac{2t}{\pt}} 
 & \frac{\pi \pt}{4 \sqrt{3}}
 & \left]-\infty,\infty\right[ 
 & 2 
 & \frac{1}{96}\Gamma^3\left(T_{\rm a}-\pt\right)^3 
 & 0
 \\ \hline
 \textit{Lorentzian} 
 &\sqrt{\frac{2}{\pi \pt}}\frac{\pt^2}{\pt^2+t^2} 
 & \pt
 & \left]-\infty,\infty\right[ 
 & \frac{23}{25} 
 & \frac{3}{20}\:\Gamma^{\frac{5}{2}}(T_{\rm a}-\pt)^{\frac{5}{2}} 
 & 0
 \\ \hline\hline
 \text{Gaussian} 
 & \frac{1}{\pi^{1/4}\sqrt{\pt}}\exp{\left(-\frac{t^2}{2\pt^2}\right)} 
 & \frac{\pt}{\sqrt{2}}
 & \left]-\infty,\infty\right[ 
 & - 
 & - 
 & \frac{\exp\left(1-\frac{9}{16 \pt^4\Gamma^4}-\frac{1}{2\pt^2\Gamma^2}-\pt^2\Gamma^2\right)}{16\sqrt{\pi}\pt^5\Gamma^5}
 \\ \hline
 \text{Increasing exp.} 
 & \frac{1}{\sqrt{\pt}}\exp{\left(\frac{t}{2\pt}\right)}
 & \pt
 & \left]-\infty,0\right] 
 & - 
 & - 
 & 1-4^{-\xi} \left( \frac{1+\xi}{1+2\xi} \right)^{-1-2\xi}, \xi=\frac{1}{2\Gamma \pt - 1}
 \\
\end{array}$
\caption{Considered set of  pulses for infinite time intervals, i.e., untruncated pulses. For each, we list its 
shape $\phi(t)$,
time variance $\ct $, domain $[t_1,t_2]$, threshold time $T_{\rm a}$ (if applicable), and asymptotic behavior of the inefficiency $1-\eta$ for either $\pt\rightarrow T_{\rm a}$ or $\pt\rightarrow \infty$. As explained in the main text, a subset of the pulses (first three, italic) have tails that do not decay faster than exponential, and therefore attain unit efficiency for $\pt \geq T_{\rm a}$. The Gaussian and the increasing exponential shapes, on the other hand, decay faster than exponential and do not have threshold times. We study their asymptotic behaviour for $\pt\rightarrow \infty$ (the one of the increasing exponential is exact).
\\
}
\label{table:inf_pulses}
\end{table*}
From an arbitrary mode shape $\phi$ and the system's characteristics (summarized in the parameter $\Gamma$), our protocol yields an optimized storage strategy. In this and the next sections, we study the  performance of the method for the pulse shapes in Table~\ref{table:inf_pulses}. Here it is important to note that the pulses can be characterized by multiple different times. First of all, there is the characteristic time of the pulse $\pt$ -- describing the width of the peak. $\pt$ is chosen to provide a simple analytical description of the shape. This, however, means that there is some arbitrariness in how $\pt$ is defined. To allow a comparison of different pulse shapes, we therefore also introduce the time variance 
\cite{Giannelli2018}
\begin{equation}\label{eq:coh_time}
    \ct =\sqrt{\langle t^2\rangle - \langle t \rangle^2},
\end{equation}
where $\langle t^\alpha\rangle = \int_{t_1}^{t_2}t^\alpha\abs{\phi(t)}^2 dt$ for $\alpha$ being a non-negative integer. 

Once a pulse shape is known we can relate $\pt$ and $\ct $, as showcased (for infinitely long pulses $t_2 - t_1 \rightarrow \infty$) in Table~\ref{table:inf_pulses}. Notice that, when $\phi$ is truncated (see Sec.~\ref{sec:finite_pulses}), $\ct $ varies depending on the truncation.
Another time scale is the time interval $t_2-t_1$ on which the pulse is defined, which we will show also has an influence on the efficiency. To understand how various effects influence the efficiency $\eta$, in this section we consider pulses over an infinitely long time interval $t_2-t_1\rightarrow \infty$. In Sec.~\ref{sec:finite_pulses}, we generalize the results to finite intervals $t_2-t_1$. As we will see, different shapes $\phi$ lead to different responses of the memory and thus varying efficiencies.

%

The analytical expressions for the efficiencies presented here and in the next Sec.~\ref{sec:finite_pulses} are derived in the high efficiency regime $\eta\approx 1$, that is most interesting for the application of quantum memories. The general procedure is the following. 
%
By plugging Eq.~\eqref{eq:Psi} into Eq.~\eqref{eq:eta_r} and using $c=1/\sqrt{\eta}$ (see Sec.~\ref{sec:optimal_memories} and App.~\ref{app:c_opt}), we can derive $\eta$ from $\phi(t)$ and $t_c$
%
%
\begin{equation}
\label{eq:fid_inf_pulse}
\begin{split}
    \eta=&
    \int_{t_1}^{t_c} \absa{\phi(t)}^2 dt
    +
    \phi^*\left(t_c\right)\int_{t_c}^{t_2}\phi(t)e^{-\Gamma\left(t-t_c\right)}dt 
    .
    \end{split}
\end{equation}
%
For any incoming pulse shape $\phi$, we can then plug Eq.~\eqref{eq:psi_inequality} (that is saturated at the critical time) 
into this last Eq.~\eqref{eq:fid_inf_pulse} to find
\begin{equation}\label{eq:crit_time_exact}
    \left\lvert \phi(t_c)\right\rvert^2 =
    2\Gamma \phi^{*}(t_c)\int_{t_c}^{t_2}\phi(t)e^{-\Gamma(t-t_c)}dt.
\end{equation}
From this exact relation $t_c$ is determined, and then employed to calculate $\eta$ via Eq.~\eqref{eq:fid_inf_pulse}. These last two steps are done perturbatively in the parameter $(\Gamma \pt)^{-1} \ll 1$, implying that the resulting expression for $\eta$ is valid in the high efficiency regime.

\begin{figure}[]
	\centering
	\includegraphics[width=\columnwidth]{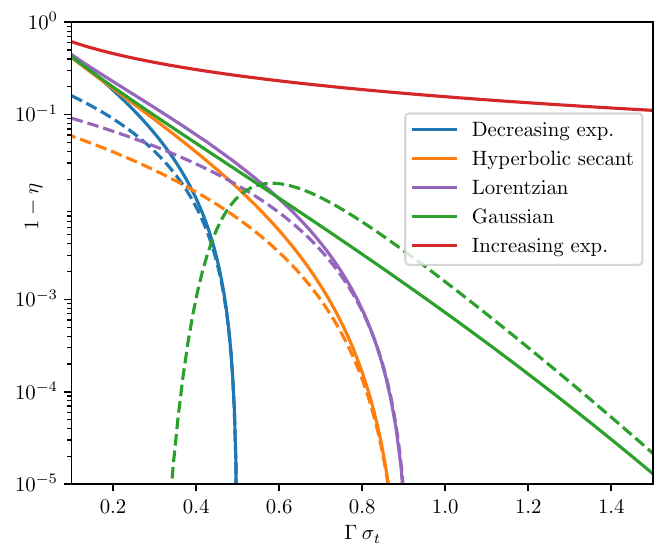}
	\caption{
Inefficiencies for different pulses defined on infinite intervals as a function of $\Gamma \ct $, where $\ct $ is the photon time variance of the pulse shape in Eq.~\eqref{eq:coh_time}. Solid and dashed lines show the optimized result from our protocol (see main text for details), and the asymptotic expansions for $\pt\rightarrow T_{\rm a}$ or $\pt\rightarrow \infty$ (see Table~\ref{table:inf_pulses}), respectively. For the increasing exponential, the efficiency expansion for $\pt\rightarrow \infty$ is exact and therefore the dashed and full lines always overlap.
}
	\label{fig:Infinite_pulses}
\end{figure}

To demonstrate the procedure, we first consider retrieval into an exponentially decreasing pulse $\phi(t)=\exp(-t/2\pt)/\sqrt{\pt}$, as defined in the first row of Table~\ref{table:inf_pulses}.
When $\pt$ is larger than a {\it threshold} time $T_{\rm a}=1/2\Gamma$, the desired outgoing shape 
decays slower then the fastest possible atomic readout, which is achieved by transferring all remaining excitations to the atomic $\ket{e}$ levels (this strategy is identified as fast storage/retrieval in Ref.~\cite{Gorshkov2007Universal}).
Since it is always possible to reduce the readout rate
by placing some of the excitations in the levels $\ket{s}$ (thus lowering $P$), we can perfectly follow the desired shape whenever $\pt \geq T_{\rm a}$ and the inequality in Eq.~\eqref{eq:psi_inequality} is never violated. 
For $\pt\geq T_{\rm a}$ the storage and retrieval efficiencies are thus ideal and $\eta =1$. 

In contrast, for $\pt<T_{\rm a}$ the desired shape decays faster than the atomic excitations from $\ket{e}$ and it is impossible to get the excitation out as fast as desired. This means that the inequality in Eq.~\eqref{eq:psi_inequality} is violated at the initial time $t_1$, implying $t_c=t_1$. Our protocol then prescribes that we should read out the memory as fast as possible by immediately transferring all excitations to the level $\ket{e}$, resulting in $\psi(t)=\sqrt{2\Gamma}\exp(-\Gamma t)$ \footnote{Note that, for the decreasing exponential, $c\neq 1/\sqrt{\eta}$. As explained in Sec.~\ref{sec:optimal_memories_psi_optimization}, since $t_1 = t_c$, $c$ is limited by the maximal retrieval rate out of the system, that is achieved via the $\psi(t)$ chosen here.}. The efficiency is then given by the overlap of the two exponentials [see Eq.~\eqref{eq:fid_inf_pulse}] and yields $\eta=8\pt\Gamma/(2\pt\Gamma+1)^2$, corresponding to a near-threshold behavior of the ineffieciency $1-\eta= \Gamma^2(T_{\rm a}-\pt)^2$ for $\pt\rightarrow T_{\rm a}^-$.


The above example illustrates a general behavior. For incoming (retrieved) pulse shapes increasing (decreasing) faster than an exponential with a rate $\Gamma$, perfect storage (retrieval)  is impossible. In the opposite scenario, perfect storage (retrieval) can always be attained and $\eta=1$. We can therefore divide the pulse shapes into two categories (as highlighted in Table~\ref{table:inf_pulses}). In the first, we have all $\phi$ with tails decaying at most exponentially. For these shapes, unit efficiency can always be achieved when their characteristic timescales $\pt$ are longer than the associated threshold times $T_{\rm a}$, i.e. $\eta=1$ for $\pt > T_{\rm a}$. In these cases, we can determine an exponent $x$ and a constant $A$ such that $\eta \approx 1-A\Gamma^x\left(\pt-T_{\rm a}\right)^x$ for $\pt$ approaching $T_{\rm a}$ from below. On the other hand, shapes $\phi$ having tails decaying faster than exponential are collected in a second category for which unit efficiency is unattainable, but we can look at the asymptotic behavior for $\pt\rightarrow \infty$. The asymptotic behaviours of the pulses are summarized in Table~\ref{table:inf_pulses}.

\begin{figure}[]
	\centering
	\includegraphics[width=\columnwidth]{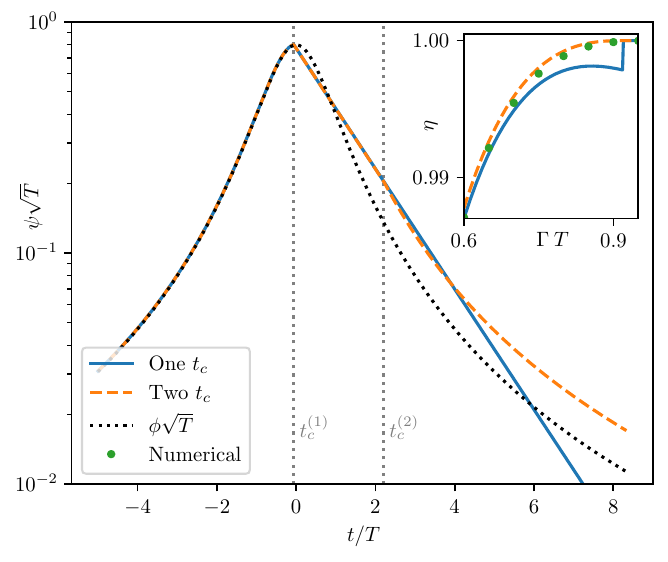}%
	\caption{
 The strategies with one and two critical times are compared to each other for a Lorentzian shape $\phi(t)$ (black dotted line, see Table~\ref{table:inf_pulses}). We show $\psi(t)$ from Eq.~\eqref{eq:Psi} (solid blue line) and Eq.~\eqref{eq:twotc} (dashed orange line), with $\Gamma \pt = 0.6$. After the second critical time $t_c^{(2)}$ (indicated alongside $t_c^{(1)}$ with  vertical grey dotted lines), the two differ. In the inset, we show the efficiency $\eta$ as a function of $\Gamma \pt$ for the two approaches and our numerical optimization (green dots). When $\pt$ is larger than $T_{\rm a}$, the inequality in Eq.~\eqref{eq:psi_inequality} is never saturated and $\eta = 1$ for both strategies. Note that for the numerical approach $\phi(t)$ must be truncated, which introduces a small inefficiency not present in the other. We employ a sufficiently large $|t_2-t_1| = 5\sqrt{3} \pi \pt$ for these truncation effects to be negligible (see Sec.~\ref{sec:finite_pulses}).}
	\label{fig:two_tc}
\end{figure}

In Fig.~\ref{fig:Infinite_pulses}, we show the inefficiencies of all pulses in Table~\ref{table:inf_pulses} as a function of $\Gamma \ct $. As it is possible to see, lines corresponding to pulses for which it is possible to identify an adiabatic time $T_{\rm a}$, namely the decreasing exponential (blue), the hyperbolic secant (orange) and the Lorentzian (violet), are characterized by vertical asymptotes at their $T_{\rm a}$. All other pulses, represented here by the increasing exponential (red) and the Gaussian (green), never attain perfect efficiencies and scale according to the limits presented in Table~\ref{table:inf_pulses}. Qualitatively, pulse shapes that are more similar to the decreasing exponential perform better. This is related to the fact that the associated map $\psi(t)$, after the critical time $t_c$, has the highest overlap with $\phi(t)$. On the other hand, the increasing exponential retains the lowest efficiency. Indeed, among all the chosen pulses, this is the one decreasing the fastest at the end of the retrieval (in fact, it is discontinuous when we abrouptly terminate the pulse at $t_2=0$), and thus the limited response time of the memory results in the worst performance. We explore this in more detail in subsection \ref{sec:finite_pulses}.

The discussion above focuses on pulses which have an exponential-like behaviour. For completeness, we finally consider  
pulses that have either more than one peak or tails that decrease slower than an exponential, e.g. the Lorentzian in Table~\ref{table:inf_pulses}.  
Resorting to one critical time $t_c$ is not optimal for these $\phi$. In this case a better strategy is to employ more critical times $t_c^{(1)},t_c^{(2)},t_c^{(3)},\dots$. This allows the map $\psi(t)$ to follow the desired shape 
$\phi(t)$ if its intensity later drops to a value that is sufficiently low to again fulfill the inequality in Eq.~\eqref{eq:psi_inequality}.
Using the map $\psi$ in Eq.~\eqref{eq:Psi} (with a single critical time), leads to a shape $\psi(t)$ that crosses the desired shape $\phi(t)$ some time after $t_c$. At that point, it is better to remove part of the excitations from $\ket{e}$ and again follow  $\phi(t)$, rather than following the exponential decay $\propto e^{-\Gamma t}$. 

This is shown in Fig.~\ref{fig:two_tc}, where we compare $\psi(t)$ with a single $t_c$ in Eq.~\eqref{eq:Psi} (full blue line) with the Lorentzian pulse $\phi(t)$ (dotted black line). As seen in the figure, the exponential decrease of $\psi(t)$ in Eq.~\eqref{eq:Psi} undershoots the desired shape at large $t$. In this case, a better strategy is to introduce two critical times such that $\psi(t)$ becomes
%
\begin{equation}
\psi(t)= \begin{cases}c\phi\left(t\right), &t<t_{c}^{(1)},\\
c\phi\left(t_{c}^{(1)}\right)e^{-\Gamma\left(t-t_{c}^{(1)}\right)}, &t_{c}^{(1)}<t<t_{c}^{(2)},\\
c\frac{\phi\left(t_{c}^{(1)}\right)}{\phi\left(t_{c}^{(2)}\right)}e^{-\Gamma\left(t_{c}^{(2)}-t_{c}^{(1)}\right)}\phi\left(t\right), &t>t_{c}^{(2)}.
\end{cases}
\label{eq:twotc}
\end{equation}
Here, the first critical time $t_{c}^{(1)}$ is found in the same way as before, while the second $t_{c}^{(2)}$ is determined such that $\int_{-\infty}^{\infty}|\psi(t)|^2 dt=1$. This ensures that all excitations in the memory are transferred to the outgoing field at the end of the retrieval process, while at the same time maximizing the overlap between $\psi(t)$ and $\phi(t)$. The shape obtained by optimizing $c$ with this $\psi(t)$ is also shown in Fig.~\ref{fig:two_tc} (orange dashed curve).

Finally, in the inset we compare the efficiencies obtained by the two analytical strategies with one [Eq.~\eqref{eq:Psi}, full line] and two [Eq.~\eqref{eq:twotc}, dashed line] critical times, respectively, as well as the OCT numerical approach (green dots). As expected, two critical times improve $\eta$ for the Lorentzian pulse when $\pt < T_{\rm a}$, where we recall that $T_{\rm a} = \Gamma^{-1} 23/25 $ for the infinite Lorentzian pulse (see Table~\ref{table:inf_pulses}). On the other hand, for $\pt \geq T_{\rm a}$ the inequality in Eq.~\eqref{eq:psi_inequality} is never violated and the efficiency is unity for both strategies. The green dots are placed in between the two curves, demonstrating that the numerical approach yields fidelities better (worse) than the strategy with one (two) critical time(s).



%
\subsection{Finite time intervals}\label{sec:finite_pulses}
In this section, we consider pulses $\phi$ that are truncated at their
final time $t_2$, i.e. $\phi(t) = 0$ whenever 
$t > t_2$ for finite $t_2$. For retrieval, the behaviour of $\phi$ at $t_1$ does not affect the memory performance (and the opposite is true for storage). We assume that the pulses are normalized on this interval, such that $\int_{t_1}^{t_2}|\phi(t)|^2 dt = 1$. In these cases, unit efficiencies are never attainable. In fact, whenever the memory is characterized by a maximal readout rate $\Gamma$, all excitations cannot be retrieved, e.g. if we excite the atoms at the initial time $t_1$,  a fraction $\exp(-2\Gamma (t_2-t_1))$ will be left at the final time $t_2$ (recall that $\Gamma$ is the decay rate for the amplitude). Equivalently, the fact that the efficiency is always below unity can also be found by noting that a pulse ending at $t=t_2$ always decays faster than an exponential (see Sec.~\ref{sec:Infinite_Pulses}). Below, we therefore determine how the truncation at 
$t_2$ influences the maximum achievable efficiency $\eta$. The situation is then similar to the increasing exponential encountered in the previous 
subsection, where we observed that the discontinuity at the final time resulted in a low efficiency. 

Since in Sec.~\ref{sec:Infinite_Pulses} we investigated how the inequality in Eq.~\eqref{eq:psi_inequality} limits the efficiency $\eta$ when it is violated within the interval $[t_1,t_2]$, we now assume $\Gamma \pt \gg 1$. In this way, we restrict ourselves to the detrimental contributions to $\eta$ coming from the edge
at $t_2$.
As before, we study the memory's behaviour in the proximity of unit efficiencies and set $c=1/\sqrt{\eta}$ in Eq.~\eqref{eq:Psi}.
%

To investigate the influence of the edges it is convenient to rewrite the efficiency as 
\begin{equation}
\label{eq:eta_s_finite}
\eta =
1-\int_{t_c}^{t_2}\left[
\left|\phi\left(t\right)\right|^2
-
\phi^*\left(t_c\right)\phi(t)e^{-\Gamma (t-t_c)}
\right]dt,
\end{equation}
where we used that the incoming pulse is normalized. 
From this equation, it is evident that $\eta$ only depends on the behavior of $\phi(t)$ between the critical and the final times, $t_c$ and $t_2$ respectively. 
Since we are looking at the regime of near-unit efficiency, it is reasonable to assume that $t_c$ is close to $ t_2$ when we are limited by edge effects. 
Therefore, we can approximate the considered pulse $\phi(t)$ for $t \in [t_c,t_2]$ by
\begin{equation}
\label{eq:Ein_finite}
    \phi\left(t\right) \approx \alpha_n\frac{1}{\sqrt{\pt}}\left(\frac{t-t_2}{\pt}\right)^n.
\end{equation}
Here, $n \geq 0$ is the power of the leading term in the Taylor series of $\phi\left(t\right)$ with respect to $(t-t_2)/\pt$, and $\alpha_n/\sqrt{\pt}$ is its coefficient. We explicitly included a factor of $1/\sqrt{\pt}$, to ensure that the coefficient $\alpha_n$ only depends on the shape and not the duration. 

Recalling again that $\phi$ is normalized and employing Eqs.~\eqref{eq:Psi} and \eqref{eq:psi_inequality}, we find
%
\begin{equation}\label{eq:E_in(t_c)^2}
\abs{\phi\left(t_c\right)}^2=2\Gamma
\left(
\eta-1+
\int_{t_c}^{t_2}\abs{\phi\left(t\right)}^2dt
\right).
\end{equation}
This last relation is then used together with Eq.~\eqref{eq:Ein_finite} to calculate 
\begin{equation}\label{eq:critical_times_trunc}
    t_c \approx t_2 - \frac{1+2n}{2\Gamma} -\frac{\pt^{2n+1}\Gamma^{2n}}{\alpha_{n}^{2}}\frac{4^n(1-\eta)}{(2n+1)^{2n-1}}
    .
\end{equation}
Plugging this last equation into Eq.~\eqref{eq:eta_s_finite} and keeping the largest contribution in the small inefficiency $1-\eta $, we finally get
\begin{equation}\label{eq:eta_s_finite_exp}
\eta = 1 - \abs{\alpha_n}^2\frac{\beta_n}{\left(\Gamma \pt\right)^{2n+1}},
\end{equation}
where the coefficients $\beta_n$ depend on the leading power $n$ of the Taylor series in Eq.~\eqref{eq:Ein_finite}. The first ones are
%
\begin{subequations}
\begin{align}
\beta_0
&=
\log 2 - \frac{1}{2}
,\\
\beta_1
&=
\frac{
9\left(
e^{\frac{3}{2}} - 4
\right)
}
{
16\left(
e^{\frac{3}{2}} - 1
\right)
}
,\\
\beta_2
&=
\frac{
625\left(
16 - e^{\frac{5}{2}}
\right)
}
{
128\left(
3 e^{\frac{5}{2}} + 2
\right)
}
,
\end{align}
\end{subequations}
while all others can be determined following the steps outlined above.

\begin{figure}[]
	\centering
	\includegraphics[width=\columnwidth]{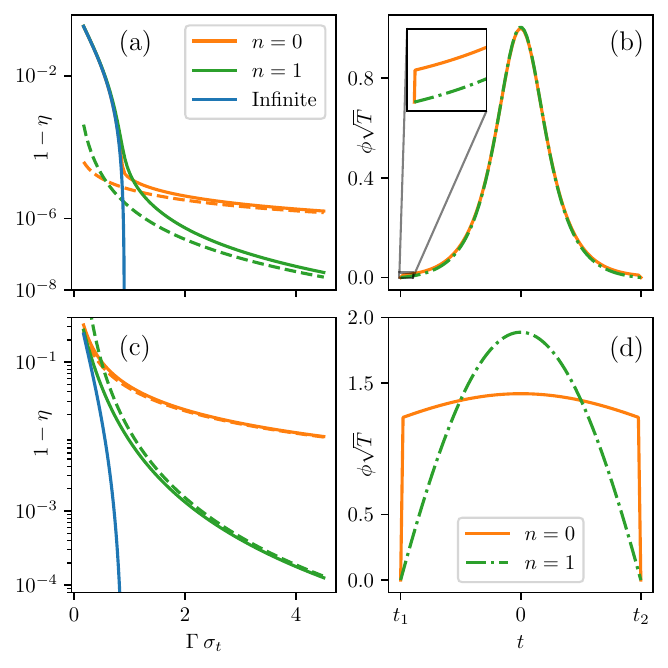}
	\caption{
 Performance of our protocol for infinitely long (blue) and truncated (orange and green) sech pulses (see Table~\ref{table:inf_pulses}). In the left column (a),(c) we present inefficiencies as functions of $\Gamma \ct $, while in the right one (b),(d) we show the associated truncated shapes. In (b,d) the full orange lines corresponds to pulses truncated by cutting off the function and renormalizing it, such that $n=0$ in Eq.~\eqref{eq:Ein_finite}. The green dashed dotted lines are also shifted to ensure $n=1$. In (a) and (c) full lines are determined via our full protocol, and dashed ones are the approximate inefficiencies that are valid in the long pulse regime and are obtained from Eq.~\eqref{eq:E_in(t_c)^2}. For (a),(b) and (c),(d) we have chosen $|t_2-t_1|$ to be equal to $\sqrt{3} \pi \pt$ and $0.1\times\sqrt{3} \pi \pt$ respectively.
 }
\label{fig:Omega_sech_expan}
\end{figure}

Eq.~\eqref{eq:eta_s_finite_exp} implies that better efficiencies are attained for pulses that are slowly decreasing between $t_c$ and $t_2$. 
In Figs.~\ref{fig:Omega_sech_expan}(a,c), we compare the inefficiencies obtained with our protocol (as described in Sec.~\ref{sec:optimal_memories} -- full lines) with Eq.~\eqref{eq:eta_s_finite_exp} (dashed lines). We use the sech pulse from Table~\ref{table:inf_pulses}, with blue, orange and green indicating the untruncated and truncated pulses with $n=0$ and $n=1$, respectively. The latter two are shown in panels (b,d). As can be  seen in the figure, when imperfections from truncation become relevant, the inefficiencies deviate from the vertical asymptotes of the infinite intervals and are well approximated by Eq.~\eqref{eq:eta_s_finite_exp}.


%
\subsection{Adiabatic elimination in the ``atom-limited" case}
\label{sec:adiabatic_elimination_bad_cavity}
In the beginning of Sec.~\ref{sec:Bad_cavity_limit} we claimed that in the ``atom-limited" regime it is possible to adiabatically eliminate the cavity mode $\mathcal{E}$ from Eqs.~\eqref{eq:eqm}. From the resulting Eqs.~\eqref{eq:gen_bc}, we then eliminated the losses to arrive at Eqs.~\eqref{eq:lossles_bc}, which we used to determine a mapping $\psi$ and the corresponding $\Omega$. In this section, we derive sufficient conditions for Eqs.~\eqref{eq:gen_bc} (when losses are present) and Eqs.~\eqref{eq:lossles_bc} (without losses) to yield the correct system dynamics. Importantly, we remark that even when the adiabatic approximation is not fulfilled, the $\Omega$ in Eq.~\eqref{eq:Omega_atom_limited_full} yields a valid (perhaps non-optimal) retrieval strategy. It is therefore always possible to validate the adiabatic elimination of $\mathcal{E}$ simply by plugging this $\Omega$ back into Eqs.~\eqref{eq:eqm}.

We recall that, following the time reversal argument in Ref.~\cite{Gorshkov2007a}, it is sufficient to consider the retrieval process such that $\mathcal{E}_{\rm in} = 0$ and therefore [from Eqs.~\eqref{eq:in-out} and \eqref{eq:map_def}] $\mathcal{E} = \psi/\sqrt{2\kappa}$. To adiabatically eliminate $\mathcal{E}$ and arrive at  Eqs.~\eqref{eq:gen_bc}, we must demonstrate that the magnitude of $\dot{\mathcal{E}}$ is negligible compared to $\kappa \mathcal{E}$ at any time $t$ of the retrieval process. This follows from Eq.~\eqref{eq:Psi}, stating that the characterizing timescales for $\psi$ and $\dot{\psi}$ are the same as for $\phi$ and $\dot{\phi}$, respectively (unless the memory is fundamentally flawed, see below). Therefore, we conclude that $\mathcal{E} = \psi/\sqrt{2\kappa} \sim 1/\sqrt{2 \kappa \pt}$ and $\dot{\mathcal{E}} = \dot{\psi}/\sqrt{2\kappa} \sim 1/\sqrt{\kappa \pt^3}$ or, in other words, $\dot{\mathcal{E}}/(\kappa \mathcal{E}) \sim (\kappa \pt)^{-1} \ll 1$, where the latter condition is the definition of the ``atom-limited" regime. 

Before investigating the opposite ``cavity-limited" regime, two remarks ought to be done. First, the conclusions above \textit{may} not hold if the memory is highly inefficient, i.e., $\Gamma \pt \ll 1$. In this case, the efficiency $\eta \ll 1$, the critical time $t_c \rightarrow t_1$ and the features of $\psi$ become very different compared to the ones of $\phi$ (see Fig.~\ref{fig:evolution} and Sec.~\ref{sec:optimal_memories_psi_optimization}). The only way to ensure that $\mathcal{E}$ can be adiabatically eliminated is a numerical investigation from the $\Omega$ prescribed by our approach (see above). The second remark relates to the characteristic timescale $\pt$ of the incoming pulse $\phi$. For clarity and simplicity, we assumed throughout this work that a single parameter $\pt$ suffices to describe $\phi$ in a given interval $[t_1,t_2]$. In practice, however, this may not be the case -- as a particular example, consider a fast oscillating field with period $T_{\rm osc}$. In this scenario, the shorter timescale $T_{\rm osc}$ should characterize the ``atom-limited" regime. In fact, $\dot{\phi} \sim 1/\sqrt{\pt}T_{\rm osc}$ and $\phi \sim 1/\sqrt{\pt}$, such that $\mathcal{E}$ can only be adiabatically eliminated when $\kappa T_{\rm osc} \gg 1$.

\section{``cavity-limited" case}
\label{sec:Good_cavity_limit}
Above, we considered memories in the ``atom-limited" case, where 
$\kappa \pt \gg 1$ and the cavity mode $\mathcal{E}$ can be adiabatically eliminated. In this section, we investigate the opposite ``cavity-limited" regime, 
where the atomic response is fast, i.e.
$\Gamma \pt \gg 1$. As we shall see, in the interesting case of efficient memories $\eta \approx 1$, it is possible to adiabatically eliminate the atomic field $P$ instead. The optimal strategy outlined in Sec.~\ref{sec:optimal_memories} for storage and retrieval can then be reutilized with minor modifications, extending our protocol to the ``cavity-limited" regime.


Contrary to the ``atom-limited" regime, where the efficiency of the quantum memory is determined by how fast the atomic ensemble couples to the optical field (the parameter $\Gamma$), the limiting factor will now  be the rate $\kappa$ at which the cavity mode $\mathcal{E}$ couples to the outside world.
In this regime, it is not possible to adiabatically eliminate $\mathcal{E}$, and Eqs.~\eqref{eq:gen_bc} cease to be valid. Unfortunately, this means that we cannot relate the lossy and the lossless cases as we did in Eqs.~\eqref{eq:efficiencies_loss_lossless} of Sec.~\ref{sec:Model}, making the analysis and subsequent optimization of the memory in the ``cavity-limited" regime more challenging. However, as demonstrated below, it is still possible to lower the system's complexity by resorting to approximations that are valid in the near-adiabatic regime.

As proven in App.~\ref{app:elimination_p} and discussed in Sec.~\ref{sec:adiabatic_elimination_good_cavity}, when the cooperativity is high $C\gg 1$ \cite{Tolazzi2021},
$P$ can be adiabatically eliminated by setting $\dot{P}=0$ and we obtain
\footnote{
Note that Eqs.~\eqref{eq:EQM} are not valid when $\gamma=\Delta=0$. In this case, the equations of motion are reduced to one differential equation: $\mathcal{E}_\text{out}=-\frac{\sqrt{2\kappa}\Omega}{g}S-\mathcal{E}_\text{in}$, $\dot{S}=-\frac{\dot{\Omega}\Omega^*+\kappa|\Omega|^2}{g^2+|\Omega|^2}S-\frac{\sqrt{2\kappa}g\Omega^*}{g^2+|\Omega|^2}\mathcal{E}_\text{in}$.
}
\begin{subequations}\label{eq:EQM}
\begin{align}
\mathcal{E}_\text{out} &= \sqrt{2\kappa_\text{in}}\mathcal{E}-\mathcal{E}_\text{in}\label{eq:EQM1},\\
\dot{\mathcal{E}} &= -\left(\kappa+\frac{g^2N}{\gamma+i\Delta}\right)\mathcal{E}-\frac{g\sqrt{N}\Omega}{\gamma+i\Delta}S+\sqrt{2\kappa_\text{in}}\mathcal{E}_\text{in} \label{eq:EQM2},\\
\dot{S} &= -\frac{\left|\Omega\right|^2}{\gamma+i\Delta}S-\frac{g\sqrt{N}\Omega^*}{\gamma+i\Delta}\mathcal{E}.
\end{align}
\end{subequations}
%
%
By comparing these equations of motion 
with the ones for the ``atom-limited" case in Eqs.~\eqref{eq:gen_bc}, a similarity and a difference are evident. 
First, we can find a similar relation between the efficiencies when the cavity losses are present and absent. 
This can be seen by rescaling the output field in Eqs.~\eqref{eq:EQM} according to  $\mathcal{E}_\text{out}\rightarrow \sqrt{\kappa/\kappa_\text{in}} \mathcal{E}_\text{out}$ during retrieval. 
Then, from Eq.~\eqref{eq:eta_r-def-gen} we find an identical factor $\kappa_\text{in}/\kappa$ that multiplies $\eta$ and the retrieval efficiency is simply reduced by a factor $\kappa_\text{in}/\kappa$. Without loss of generality, we therefore set $\kappa_\text{loss} = 0$ and $\kappa=\kappa_\text{in}$ in the following, noting that cavity losses can always be added later.

Secondly, the difference
relates to the losses from spontaneous emission. Since the timescale at which excitations are transferred from the atomic ground $\ket{g}$ to the storage $\ket{s}$ states (via $\ket{e}$) can be short compared to the lifetime $\kappa^{-1}$ of the optical field $\mathcal{E}$, there is no longer a one-to-one correspondance between the cavity field $\mathcal{E}$ and the atomic polarization  $P$. This implies that the decay from the excited atomic states $\ket{e}$ has a non-trivial relation to $\mathcal{E}$ and the outgoing field $\mathcal{E}_\text{out}$. As a result, depending on the desired shape $\phi$ the detrimental contributions from the decay $\gamma$ can be vastly different. Therefore, it is not possible to find a correspondence between the lossless and the lossy cases, as we  did in Eqs.~\eqref{eq:efficiencies_loss_lossless}. It is, however,  still possible to find an upper bound on the losses due to the atomic decay $\gamma$, and hence understand its detrimental effect on the efficiency. As we demonstrate in App.~\ref{app:elimination_p}, for the considered strategy and $\kappa \pt\gtrsim 1$ the probability $P_\gamma$ to emit a photon during the process can be estimated to be 
\begin{equation}
   P_\gamma = 
   2\gamma \int_{t_1}^{t_2} \left \lvert P(t') \right\rvert^2 dt' 
   \sim
   \frac{1}{C}.
\end{equation}
%
This last equation tells us that, following our optimal strategy, the excitations lost due to atomic decay are determined by the inverse cooperativity $C$. Therefore, albeit not being able to exactly relate the lossy and lossless case to each other, we can estimate the efficiency $\eta$ from
\begin{equation}
    \eta \approx 
    \frac{\kappa_{\rm in}}{\kappa} 
    \left( 
    \eta^{(0)} -P_\gamma 
    \right),
\end{equation}
with $\eta^{(0)}$ being the lossless efficiency (of storage or retrieval, equivalently) and $P_\gamma\sim 1/C$ for reasonable pulse durations $\kappa \pt\gtrsim 1$.

Since the description of the memory in terms of the mappings $\psi_s$ and $\psi_r$ is guaranteed by the linearity of the full equations of motion in Eqs.~\eqref{eq:eqm}, we can utilize it below to generalize our protocol to the ``cavity-limited" regime 
$\Gamma  \pt \gg 1$.
In fact, the only difference compared to the ``atom-limited" regime is that Eq.~\eqref{eq:conservation}, describing the conservation of excitations in the system, has to be modified to
\begin{equation}\label{eq:total_excitations_good_cavity}
\abs{S(t)}^2+\abs{\mathcal{E}(t)}^2+\int_{t_1}^t\abs{\mathcal{E}_\text{out}(t')}^2dt'=\abs{S(t_1)}^2
\end{equation}
for all $t$. Here, the long cavity lifetime $\kappa^{-1}$ obliges us to include $\mathcal{E}$ in the equation, and $|P|^2$ has been neglected following its adiabatic elimination. 


%
Resorting to Eqs.~\eqref{eq:E_out_psi_S} and \eqref{eq:EQM1} to express the functions $\mathcal{E}$ and $\mathcal{E}_\text{out}$ in terms of $\psi$, Eq.~\eqref{eq:total_excitations_good_cavity} can be rewritten as
\begin{equation}
\frac{\abs{S(t)}^2}{\abs{S(t_1)}^2}+\frac{1}{2\kappa}\abs{\psi(t)}^2+\int_{t_1}^t\abs{\psi(t')}^2dt'=1,
\label{eq:total_excitations_good_cavity_2}
\end{equation}
for all $t$. From this equation, we find the condition that $\psi$ must satisfy in the ``cavity-limited" case
\begin{equation}\label{eq:psi_inequality_good_cavity}
\abs{\psi(t)}^2\leq 2\kappa\left(1-\int_{t_1}^t\abs{\psi(t')}^2dt'\right).
\end{equation}
As anticipated above, by comparing this equation with the corresponding expression for the ``atom-limited" regime~\eqref{eq:psi_inequality}, we note that here the limiting rate is represented by $\kappa$ rather than $\Gamma$. Besides that, there is no qualitative difference between Eqs.~\eqref{eq:psi_inequality} and \eqref{eq:psi_inequality_good_cavity}. Therefore, the arguments in Sec.~\ref{sec:optimal_memories} that allowed us to identify the optimal mapping $\psi(t)$ in Eq.~\eqref{eq:Psi} are still valid here, provided we substitute $\Gamma$ with $\kappa$. In other words, the optimal mode function $\psi$ for storage or retrieval in the ``cavity-limited'' case is the same as in the ``atom-limited'', only with $\kappa$ and $\Gamma$ interchanged. This includes identifying the critical time(s) $t_c$ and the optimization over the parameter $c$ in Eq.~\eqref{eq:Psi} to find the best efficiencies for a given input pulse $\phi$. Therefore, in the ``cavity-limited'' regime we find an identical behaviour as the one in Fig.~\ref{fig:Optimization}(a), only with $\Gamma$ exchanged by $\kappa$.

The above argument demonstrates that the limitations of the ``cavity" and ``atom-limited" cases are mathematically the same, apart from the difference in which decay rate, $\kappa$ or $\Gamma$, to consider. The only other (experimentally) relevant difference is the shape of $\Omega(t)$ in the ``cavity-limited" case. Similarly to the ``atom-limited" regime considered in Sec.~\ref{sec:omega_determination}, we again solve for the field in terms of  the desired solution $\psi$. 
We isolate $\Omega(t)$ in Eq.~\eqref{eq:EQM2} with $\mathcal{E}_\text{in}=0$ and plug in $\abs{S}$ from Eq.~\eqref{eq:total_excitations_good_cavity_2} together with $\mathcal{E}=\psi S(t_1)/\sqrt{2\kappa}$ to get
%
\begin{equation}
\label{eq:Omega_good_cav}
\Omega(t)=
\begin{cases}
-\frac{
\left(\gamma+i\Delta\right)\dot{\psi}(t)+\left[\left(\gamma+i\Delta\right)\kappa +g^2N\right]\psi(t)
}{
g\sqrt{N}\sqrt{2\kappa-\abs{\psi(t)}^2-2\kappa\int_{t_1}^t\abs{\psi(t')}^2dt'}
}
e^{-i\theta(t)}, & t < t_c,\\
\gg g\sqrt{N}, & t \geq t_c.
\end{cases}
\end{equation}
The phase of $S$ defined by $\theta$ is found with the step-by-step procedure presented in App.~\ref{app:Phase_S}.
Fig.~\ref{fig:Crossover}(c-d) demonstrates that the optimal driving fields $\Omega(t)$ differ in the two limits. However, with the same values of $\Gamma$ and $\kappa$ in the ``atom" and ``cavity-limited" cases, respectively, these $\Omega(t)$ yield the exact same shape of the output field and therefore the same efficiency.

The form of 
$\Omega(t)$ in Eq.~\eqref{eq:Omega_good_cav} is well defined as long as $\abs{S(t)} \neq 0$, i.e., when $t \leq t_c$. For $t \overset{+}{\rightarrow} t_c$ we find a similar behaviour as in the ``atom-limited" regime (see Eq.~\eqref{eq:Omega_st1} and discussion below), where 
$\Omega(t_c) \overset{+}{\rightarrow} \infty$. 
The detailed behavior of the divergence  in 
$\Omega(t)$ near $t_c$ is different in the ``cavity-limited'' regime. 
After the critical time we want to retrieve the excitations as fast as possible by putting all population in the cavity mode.  
Since the cavity is (near) resonant with the atoms, excitations will oscillate between the cavity $\mathcal{E}$ and the atom(s) for $\Omega(t) = 0$, thus lowering the population in the cavity and the rate at which $\mathcal{E}_\text{out}$ is generated. 
To achieve the fastest retrieval and the associated exponential decay $\propto e^{-\kappa (t-t_c)}$ of $\mathcal{E}_\text{out}$, we must prevent the fields $\mathcal{E}$ and $P$ from  exchanging excitations. To do this, we decouple $\ket{e}$ from $\ket{s}$ by turning up the magnitude of $\Omega(t)$ for $t>t_c$. 

This strategy is best explained in the dressed state picture $\ket{\pm}=(\ket{e}\pm\ket{s})/\sqrt{2}$ \cite{Lukin2006}, where it is possible to show that the energy separation of $\ket{+}$ and $\ket{-}$ is $2|\Omega|$ (assuming for simplicity $\Delta = 0$). Therefore, by setting $\Omega \gg g\sqrt{N}$ for $t>t_c$, $\mathcal{E}$ is decoupled from the atom(s) and we obtain the desired exponential decay $\mathcal{E}_\text{out}\propto e^{-\kappa (t-t_c)}$ of the excitations in the cavity mode. This can be seen in Fig.~\ref{fig:Crossover}(c-d), where the orange line, corresponding to $\Omega$ in the ``cavity-limited" regime, is not set to zero after $t_c$ (as for the blue line), but diverges/is increased to sufficiently high values.

\subsection{Adiabatic elimination in the ``cavity-limited" case}
\label{sec:adiabatic_elimination_good_cavity}
In this section, we derive conditions that are sufficient (but not necessary) for adiabatically eliminating field $P$ in the ``cavity-limited" case $\Gamma \pt \gg 1$. Importantly, as for the opposite case investigated in Sec.~\ref{sec:adiabatic_elimination_bad_cavity}, our approach always yields a feasible strategy from which it is possible to check whether $\dot{P}$ is negligible or not. Furthermore, if pulse $\phi$ has multiple different timescales, the shortest one must characterize parameter $\pt$ in the defining limit $\Gamma \pt \gg 1$. Finally, the time-reversal argument can be employed again to restrict the following discussion to the retrieval process and subsequently write $\mathcal{E} = \psi/\sqrt{2\kappa}$.

A sufficient condition for adiabatically eliminating $P$ is that $\dot{P} \ll g \sqrt{N}\mathcal{E}$ in Eq.~\eqref{eq:eqm2}. By differentiating both sides of Eq.~\eqref{eq:eqm1} we can estimate the magnitude of $\dot{P}$ to be
\begin{equation}
\label{eq:adiab_el_cav}
    \frac{\dot{P}}{g\sqrt{N} \mathcal{E}}  \sim \frac{1}{\Gamma \pt} \left( 1 + \frac{1}{\kappa \pt} \right).
\end{equation}
Here, we used that the characteristic timescale of $\psi$ is not smaller than that of $\phi$ (and derivatives accordingly) and have replaced derivatives by $1/\pt$. We are mainly interested in memories, which have a not too low efficiency.  As argued in
Sec.~\ref{sec:adiabatic_elimination_bad_cavity},  in the ``cavity-limited" case this means that $\kappa \pt \gtrsim 1$. This automatically implies that the right hand side in Eq.~\eqref{eq:adiab_el_cav} is negligible for $\Gamma \pt\gg 1$, verifying the validity of the adiabatic elimination of $P$.

\section{Connecting the ``cavity'' and the ``atom-limited'' cases}
\label{sec:connection_regimes}
Above, we have investigated the memory behavior in the ``cavity-limited'' and ``atom-limited'' cases. In this last section, we test our approach by comparing our results to the numerically optimized efficiencies from Ref.~\cite{Giannelli2018}, addressing varying coupling strengths $g$. Here, $g$ is varied to switch between the two regimes. This allows us to confirm that in the limits of small and large $g$, respectively, our protocol does yield a (near) optimal result. At the same time it will shed light on the intermediate regime where the adiabatic approximations at the basis of our strategies may break down.
\begin{figure}[]
	\centering
    \subfigure{\includegraphics[width=\columnwidth]{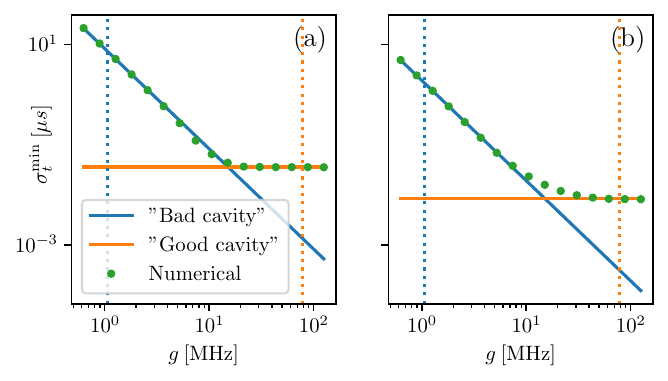}}
	\subfigure{\includegraphics[width=\columnwidth]{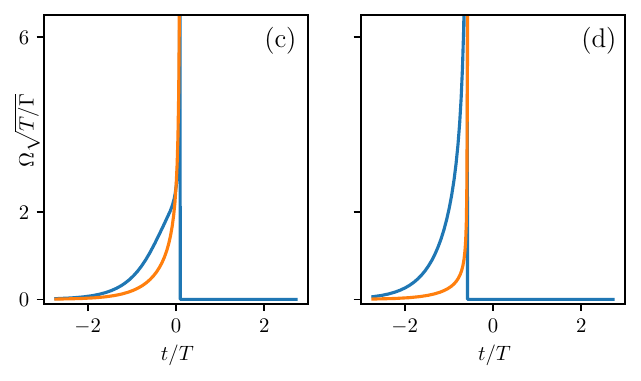}}
	\caption{
(a),(b): Minimal pulse durations $\ct ^{\rm min}$ for storage with a predetermined minimal efficiency (see main text) resulting from the numerical approach in Ref.~\cite{Giannelli2018} (green dots) and our protocol (full lines). As a testbed, we use the sech pulse (see Table~\ref{table:inf_pulses}) truncated and renormalized such that $|t_2-t_1| = \sqrt{3} \pi \pt$ and $\int_{t_1}^{t_2}|\phi|^2(t) dt = 1$. In panels (a),(c) and (b),(d) the threshold efficiencies employed for determining $\ct ^{\rm min}$ are $0.99$ and $2/3$, respectively, for $\kappa=2.42\times 2\pi\:\text{MHz}$ and $N=1$. In both cases, the solid lines cross at $\kappa=\Gamma$ due to the similarity of the two regimes.
(c),(d): Shapes of $\Omega(t)$ in the ``atom-limited" case (blue) and the ``cavity-limited" (orange) associated to the values of $g$ highlighted by the vertical dotted lines in (a),(b). The plot is normalized such that the magnitude is comparable to $\mathcal{E}_\text{in}(t)$ in each of the limits.}
	\label{fig:Crossover}
\end{figure}

When $g$ is varied, the bandwidth of the memory changes (e.g., $\Gamma$ increases with $g$ in the ``atom-limited'' regime). Rather than considering the efficiency $\eta$ for a certain pulse duration, we therefore consider the minimum time variance $\ct ^{\rm min}$ required to obtain a desired efficiency.
We choose $\ct ^{\rm min}$ as a direct measure of the shortest pulse length that can be stored in the memory 
and for comparison with the results presented in Ref.~\cite{Giannelli2018}. In that work, 
a numerical procedure was used for maximizing the efficiency. As we will see, our results are in close agreement, which again suggests that both approaches are (near) optimal in their respective regimes. 

Let us first give a qualitative description of the impact of varying $g$. When $g \ll \kappa /\sqrt{N}$ the atomic bandwidth is smaller than the cavity one. Therefore, provided $\pt$ is sufficiently large, we are in the ``atom-limited'' case
and the cavity field can be adiabatically eliminated (see Sec.~\ref{sec:adiabatic_elimination_bad_cavity}). 
The readout rate is then characterized by $\Gamma = Ng^2/\kappa$. Since $1/\Gamma$ fully characterizes the fastest possible response time of the system, it follows that $\ct ^{\rm min} \propto g^{-2}$. In the ``atom-limited'' regime, a larger coupling strength $g$ thus leads to a smaller $\ct ^{\rm min}$.

In the opposite limit $g \gg \kappa/\sqrt{N}$ the photons interact strongly with the atoms, and the memory is limited by how fast the incoming pulse can enter or leave the cavity -- i.e., we  are in the ``cavity-limited'' regime. As long as the incoming photon can enter the cavity, it can be stored. Equivalently, whenever the target shape varies on a timescale that is longer than the cavity lifetime, the stored excitation can be retrieved into it (see Sec.~\ref{sec:adiabatic_elimination_good_cavity}). The system response is thus fully characterized by parameter $\kappa$, implying that in the ``cavity-limited'' regime $\ct ^{\rm min}$ is independent of $g$.

These qualitative arguments are numerically verified in Fig.~\ref{fig:Crossover}(a-b), where the green dots are from Ref.~\cite{Giannelli2018} and the solid lines are obtained via our protocol. In the figure, we consider the sech pulse (see Table~\ref{table:inf_pulses}) and set the efficiency threshold for $\ct ^{\rm min}$ to $0.99$ [panels (a) and (c)] and $2/3$ [panels (b) and (d)]. Here, we find the solid curves by reading off the values $b_{99\%} \approx 0.549$ and $b_{2/3} \approx 0.128$ for $\Gamma \ct $, which respectively give $\eta=99\%$ and $\eta=2/3$ in Fig.~\ref{fig:Infinite_pulses} [note that edge effects are negligible in the settings investigated here -- cf. Fig.~\ref{fig:Omega_sech_expan}(a)]. From here, the minimal time variance is $\ct ^{\rm min}=b_{99\%}/\Gamma$ ($\ct ^{\rm min}=b_{2/3}/\Gamma$) and $\ct ^{\rm min}=b_{99\%}/\kappa$ ($\ct ^{\rm min}=b_{2/3}/\kappa$) in the ``cavity'' and ``atom-limited'' cases, respectively.

As seen in Figs.~\ref{fig:Crossover}(a-b), the agreement between our method and the numerical results from Ref.~\cite{Giannelli2018} is excellent both for small and large values of $g$. Furthermore, even though our result are -- strictly speaking -- not valid in the regime $g\sqrt{N}\sim \kappa$ the results obtained in the two limits do provide a rough understanding of the pulse duration in this crossover regime also. 

%
\section{Conclusion}\label{sec:conclusions}
We have investigated the performance of a cavity based $\Lambda$-type quantum memory in the nonadiabatic regime, where the memory efficiency is limited by the finite bandwidth of the memory. After simplifying the equations of motion in the ``atom'' and ``cavity-limited'' regimes we identified that in each case the bandwidth of the memory is limited by a single rate, namely, the effective atomic decay $\Gamma=Ng^2/\kappa$ and the cavity decay rate $\kappa$, respectively. Based on this we devised a strategy for maximizing the efficiency. 

Apart from the limitation imposed by having a finite decay rate, our strategy is based on the observation that the shape of a retrieved photon can be varied at will by a suitable choice of the control field $\Omega(t)$. We can thus always find the required control field $\Omega(t)$ \textit{after} the ideal shape $\psi$ has been identified. Mathematically, this yields a much simpler problem than the conventional strategy of directly optimizing the control field $\Omega(t)$. In particular, we can directly give a physically motivated ansatz for the optimal shape $\psi$. While we cannot prove that this particular solution is truly optimal, in all investigated cases our strategy yielded better solutions than  found numerically. As can be seen from Fig.~\ref{fig:Optimization}(b-e), OCT generates control shapes $\Omega(t)$ that closely resemble those found by our method. Deviations in regions where $\Omega$ diverges explain why the efficiencies we find are always better.


Because of the simplicity of our procedure, we have been able to derive analytical results for the memory behavior. Once we identified a single critical parameter for the ``atom-limited'' (``cavity-limited'') regime, i.e. $\Gamma$ ($\kappa$), the memory response $\psi$ can be characterized solely by the shape of the wave packet $\phi$ and the product of the bandwidth and pulse duration $\Gamma \pt$ ($\kappa \pt$). From $\psi$, one can then determine experimentally relevant quantities such as the optimal efficiency $\eta$ and the control shape $\Omega$. The latter will generally be different in the two regimes despite $\eta$ and $\phi$ being the same.

For pulses defined on an infinite time interval, the storage (retrieval) efficiency depends on how the early (late) tail of the pulse $\phi$ approaches zero. If it is at most exponential, ideal performance is achieved once the pulse duration is larger then a certain threshold time $T_{\rm a}$. 
%
On the other hand, pulses defined on finite intervals can only approach unit efficiency asymptotically. The efficiency $\eta$ depends on the cutoff at the beginning (end) of the pulse for storage (retrieval). Higher-order transitions lead to a slower decrease in efficiency, i.e., better performance is reached when $\phi$ is smoothly varying, so that we are closer to adiabaticity.

We note that while our theory provides a solid understanding of cavity-based quantum memories, there are multiple other platforms \cite{Bienfait2019Phonon,Axline2018Demand,palomaki2013coherent,stute2012tunable,lvovsky2009optical,Schmit2012Retrieval,vepsalainen_quantum_2016} that do not conform to the considered framework. Several experiments do not rely on optical cavities, but use atomic ensembles in free space \cite{Hammerer2010Quantum,eisaman2005electromagnetically,Cho2016Coherent,zhao2009long,Reim2011Single,tey2008strong,Hsiao2018High}. At the same time, multiple approaches use various types of inhomogeneous broadening \cite{Kraus2006Quantum,clausen2011quantum,Afzelius2010Impedance}. Such memories have been argued to be particularly useful for creating multimode quantum memories \cite{Afzelius2009Multimode,Teja2019Photonic,Sinclair2014Spectral}. Although our work does not directly apply to these situations, the techniques that we develop may be adapted to optimize the performance of these other memories. 

\section*{Acknowledgements}
We thank Arianne Brooks for her help with Fig.~\ref{fig:raman}, and  Oxana Mishina, Klaus Mølmer, and Christiane Koch for useful discussions. We acknowledge the support of Danmarks Grundforskningsfond (DNRF 139, Hy-Q Center for Hybrid Quantum Networks). T.S. and G.M. are funded by the Deutsche Forschungsgemeinschaft (DFG, German Research Foundation) -- Project-ID 429529648 -- TRR306 QuCoLiMa (``Quantum Cooperativity of Light and Matter''), and by the German Ministry of Education and Research (BMBF) via the Project NiQ (Noise in Quantum Algorithms). L.G. acknowledges the QuantERA grant SiUCs (Grant No.\ 731473) and the PNRR MUR project PE0000023-NQSTI. L.D. acknowledges the EPSRC quantum career development grant EP/W028301/1.

\clearpage
\newpage
\appendix

\section{Optimal retrieval}
\label{app:c_opt}
In this appendix, we determine the optimal value of $c$ for the case where we have a single $t_c$. Employing Eq.~\eqref{eq:Psi}, that defines the optimal $\psi(t)$, the efficiency is found from Eq.~\eqref{eq:eta_r} to be
\begin{equation}
\label{eq:eta_s_opt_c}
\begin{split}
\eta = &
\left|c\int_{t_1}^{t_c} \absa{\phi(t)}^2 dt\right.
\\
&\left.
+c\:\phi^*\left(t_c\right)\int_{t_c}^{t_2}e^{-\Gamma\left(t-t_c\right)}\phi\left(t\right)dt\right|^2.
\end{split}
\end{equation}
From the freedom in choosing $\psi(t)$ and since the constraint in Eq.~\eqref{eq:psi_inequality} does not involve the phase of $\psi(t)$, the efficiency $\eta$ must be independent of the phase of $\phi(t)$. Eq.~\eqref{eq:eta_s_opt_c} can therefore be rewritten as\tom{}
\begin{equation}
\begin{split}
\eta=& 
\left(c\int_{t_1}^{t_c} \abs{\phi(t)}^2 dt\right.
\\
&\left.
+c\:\abs{\phi\left(t_c\right)}\int_{t_c}^{t_2}e^{-\Gamma\left(t-t_c\right)}\abs{\phi\left(t\right)}dt\right)^2,
\end{split}
\end{equation}
which can be differentiated with respect to $c$ to find the minimum of $\eta$. Specifically, by taking the derivative and noting that $t_c$ also depends on $c$ -- see Eqs.~\eqref{eq:psi_inequality} and \eqref{eq:Psi} -- we find
\begin{equation}
\label{eq:eta_derivative}
\begin{split}
\pd{\eta}{c} 
= 
2\sqrt{\eta}\left(\frac{\sqrt{\eta}}{c}
-\frac{2\Gamma}{c^2\abs{\phi\left(t_c\right)}}\int_{t_c}^{t_2}e^{-\Gamma\left(t-t_c\right)}\abs{\phi\left(t\right)}dt\right),
\end{split}
\end{equation}
that, once set equal to zero, implies
\begin{equation}
    c=\frac{2\Gamma}{\sqrt{\eta}\:\abs{\phi(t_c)}}\int_{t_c}^{t_2}e^{-\Gamma\left(t-t_c\right)}\abs{\phi\left(t\right)}dt.
\end{equation}
The integral in this last relation can be calculated by expressing it in terms of $\eta$. The resulting equation
\begin{equation}
c = \frac{2\Gamma}{c\:\sqrt{\eta}\:\absa{\phi\left(t_c\right)}^2}\left(\sqrt{\eta}+\frac{c\absa{\phi(t_c)}^2}{2\Gamma}-\frac{1}{c}\right)
\end{equation}
has the solutions $c=1/\sqrt{\eta}$ and $\pm \sqrt{2\Gamma}/\absa{\phi(t_c)}$, with the first corresponding to the desired stationary point.

\section{Comparison to Refs.~\cite{Vasilev2010,Utsugi2022,Tissot2024}}
\label{app:comparison}

\begin{figure}
	\centering
	\includegraphics[width=\columnwidth]{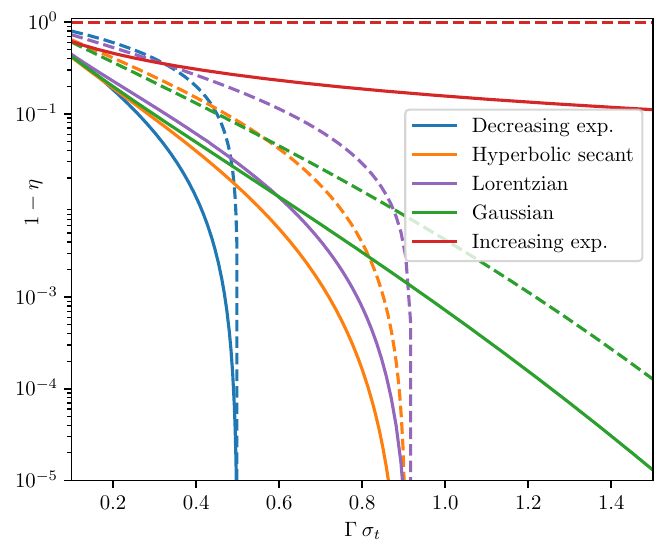}
	\caption{
Comparison between our method (full lines) and the protocols in Refs.~\cite{Vasilev2010,Utsugi2022,Tissot2024} (dashed lines). The considered settings are the same as in Fig.~\ref{fig:Infinite_pulses}.
}
	\label{fig:Methods_comparison}
\end{figure}

As explained in the main text, Refs.~\cite{Vasilev2010,Utsugi2022,Tissot2024} considered a similar strategy, but contrary to our choice $c=1/\sqrt{\eta}\geq 1$ assume a lower value $c\leq 1$ to avoid violating the inequality in Eq.~\eqref{eq:psi_inequality}. In Fig.~\ref{fig:Methods_comparison}, we compare the results obtain with our method with the results obtained with the method of Refs.~\cite{Vasilev2010,Utsugi2022,Tissot2024}. As shown in the figure, our protocol yields inefficiencies that can be orders of magnitude smaller. We note, however, that in addition to the efficiency to store an incoming mode or equivalently retrieve into a specific mode, which is our main interest here, there could also be other figures of merit. In particular our method retrieves into a (distorted) mode which has the larges possible overlap with a particular mode shape. In contrast, the methods of Refs.~\cite{Vasilev2010,Utsugi2022,Tissot2024} only retrieves into the desired mode with zero population of other modes. Which of these two characteristics is more desirable depends on the particular application. 

\section{Phase of $S$ and $\Omega$}
\label{app:Phase_S}
In Eqs.~\eqref{eq:Omega_st1} and \eqref{eq:Omega_good_cav} we give expressions for the optimal driving $\Omega(t)$ in terms of the phase $\theta$ of $S$.
Here, we derive this phase in both the ``atom" and ``cavity-limited" cases according to the definition
\begin{equation}
    S=\abs{S}e^{i\theta}.
\end{equation}
Differentiating the latter allows us to find $\dot{S}/S=\abs{\dot{S}}/\abs{S}+i\dot{\theta}$ such that
\begin{equation}
    \dot{\theta}=\text{Im}\left[\frac{\dot{S}}{S}\right].
    \label{eq:theta_dot_1}
\end{equation}
This relation can now be used to find $\theta$ in both limits.

\subsection{``atom-limited" case}
Setting $\mathcal{E}_\text{in}=0$, from Eqs.~\eqref{eq:lossles_bc} we obtain the the equations of motion in the ``atom-limited" case during retrieval:
\begin{subequations}
\begin{align}
    \mathcal{E}_\text{out}&=i\sqrt{2\Gamma}P\\
    \dot{P}&=-\left(\Gamma+i\Delta\right)P+i\Omega S \label{eq:P_retr_noloss}\\
    \dot{S}&=i\Omega^*P
    \label{eq:S_retr_noloss}
\end{align}
\end{subequations}
Inserting $S$ and $\dot{S}$ from the equations of motion into Eq.~\eqref{eq:theta_dot_1} we find
\begin{equation}
    \dot{\theta}=\frac{\abs{\Omega}}{\abs{S}}\text{Re}\left[Pe^{-i\left(\nu+\theta\right)}\right],
    \label{eq:theta_dot_2}
\end{equation}
 where $\nu$ is the phase of $\Omega=|\Omega|\exp(i\nu)$. Furthermore multiplying the complex conjugate of  Eq.~\eqref{eq:P_retr_noloss} by $P$ we find
\begin{equation}
    P e^{-i\left(\nu+\theta\right)}=\frac{i}{\abs{\Omega}\abs{S}}\left[P\dot{P}^*+\left(\Gamma-i\Delta\right)\abs{P}^2\right].
\end{equation}
Taking the real part of this equation and integrating Eq.~\eqref{eq:theta_dot_2} gives the phase
\begin{equation}
    \theta(t)=-\int_{t_1}^{t}\frac{\text{Im}\left[P(t')\dot{P}^*(t')\right]}{\abs{S(t')}^2}dt'+\int_{t_1}^{t}\frac{\Delta\abs{P(t')}^2}{\abs{S(t')}^2}dt',
\end{equation}
which can then be expressed in terms of $\psi$ using $P=\psi S(t_1)/i\sqrt{2\Gamma}$ and Eq.~\eqref{eq:S_psi_inequality}:
\begin{equation}    
\begin{split}
    \theta(t)=&-\int_{t_1}^{t}\frac{\text{Im}\left[\psi(t')\dot{\psi}^*(t')\right]}{2\Gamma-\abs{\psi(t')}^2-2\Gamma\int_{t_1}^{t'}\abs{\psi(t'')}^2dt''}dt'
    \\
    &+\int_{t_1}^{t}\frac{\Delta\abs{\psi(t')}^2}{2\Gamma-\abs{\psi(t')}^2-2\Gamma\int_{t_1}^{t'}\abs{\psi(t'')}^2dt''}dt'.
\end{split}
\end{equation}

\subsection{``cavity-limited'' case}
In the good cavity limit we employ Eq.~\eqref{eq:theta_dot_1} again, but in this case apply the equations of motion in Eqs.~\eqref{eq:EQM} with $\mathcal{E}_{\rm in}=0$. Following the same steps as in the ``atom-limited'' case with $P$ and $\mathcal{E}$ interchanged then leads to
\begin{equation}
\begin{split}
    \theta(t) = -\int_{t_1}^{t} 
    &
    \frac{\Delta}{g^2\abs{S(t')}^2}\left|\dot{\mathcal{E}}(t')+\left(\kappa+\frac{g^2}{\gamma+i\Delta}\right)\mathcal{E}(t')\right|^2
    \\
    &
    -\Delta\frac{\abs{\mathcal{E}(t')}^2}{\abs{S(t')}^2}\frac{g^2N+2\gamma\kappa}{\gamma^2+\Delta^2}
    \\
    &
    +\mathrm{Im}\left\lbrace\frac{\mathcal{E}(t')\dot{\mathcal{E}}^*(t')}{\abs{S(t')}^2}\frac{\gamma-i\Delta}{\gamma+i\Delta}\right\rbrace
    dt',
\end{split}
\end{equation}
which then can be expressed in terms of $\psi$ via Eq.~\eqref{eq:total_excitations_good_cavity_2} and using that during retrieval $\psi = \sqrt{2\kappa}\mathcal{E}$.

\section{
Losses from $\gamma$ in the ``cavity-limited'' case
}
\label{app:elimination_p}
In this appendix we show that, in the ``cavity-limited'' regime, the losses due to spontaneous emission $\gamma$ are small if we have a high cooperativity $C$ and a reasonably efficient memory. 
Let us consider the equation of motion for $P$ during retrieval ($\mathcal{E}_\text{in}=0$) and express it in terms of $\psi=\sqrt{2\kappa}\mathcal{E}$. By ignoring noise terms in Eq.~\eqref{eq:eqm1} we have that
\begin{subequations}\label{eq:eqmapp}
\begin{align}
\dot\psi=&-\kappa\psi+ig\sqrt{2N\kappa}P,
\label{eq:eqm1app}
\end{align}
\end{subequations}
so that $P$ can be rewritten as 
\begin{equation}
P=\frac{1}{g\sqrt{2N\kappa}}\left(\dot\psi+\kappa\psi\right).
\label{eq:P_good_cavity}
\end{equation}
The probability of loss due to spontaneous emission is then given by 
\begin{equation}
\label{eq:decay_p_app}
\begin{split}
    P_\gamma & = 
    2\gamma\int_{t_1}^{t_2}\left|P(t')\right|^2dt'
    \\ & =
    \frac{1}{C}\int_{t_1}^{t_2}\left|\frac{\dot{\psi}(t')}{\kappa}+\psi(t')\right|^2dt',
\end{split}
\end{equation}
where we remark that in this context $\psi$ can also be interpreted as the outgoing field. Even if we want to retrieve the stored excitation into a rapidly decaying mode, we typically have $\dot \psi \sim \kappa \psi$ (since the decay is limited by $\kappa$). Hence, the first term in the norm within the rhs of Eq.~\eqref{eq:decay_p_app} will be comparable to the second. The situation may differ if we consider input shapes $\phi$ that are rapidly oscillating at a timescale $T_\text{osc} \ll \pt$, such that the first term can attain values $\dot \psi/\kappa \sim \psi/\kappa T_\text{osc}$.
As discussed in Sec.~\ref{sec:adiabatic_elimination_good_cavity}, in this scenario we characterise the ``cavity-limited'' regime by $\Gamma T_{\rm osc} \gg 1$ and require the memory to attain reasonable efficiencies $\eta \gtrsim 1/2$. This latter assumption implies $\kappa T_{\rm osc} \gtrsim  1$ and therefore ensures $\dot \psi/\kappa \sim \psi$ in Eq.~\eqref{eq:decay_p_app}. We can therefore estimate the losses to be
\begin{equation}
    P_\gamma \sim 
    \frac{1}{C}\int_{t_1}^{t_2}\left|\psi(t')\right|^2dt'=\frac{1}{C},
\end{equation}
where we have used that $\psi$ is normalised. Hence, unless we operate in a regime with a low memory efficiency, the atom-induced losses will be roughly $P_\gamma \sim 1/C$ as in the ``atom-limited'' regime. 
We note that this discussion is analogous to the condition for the validity of adiabatically eliminating $P$ in Sec.~\ref{sec:adiabatic_elimination_good_cavity}.
\section{Numerical investigations for the exponentially decreasing pulse shape}
\label{app:decr_exp}
In this appendix, we provide details about additional numerical investigations of the decreasing exponential. As can be seen in Fig.~\ref{fig:Optimization}, this is the only shape for which the numerical method yields fidelities that are much worse than those from our approach. Below, we first qualitatively explain this discrepancy, and then investigate different settings of our numerical method to better understand its limitations.

As explained in Sec.~\ref{sec:Infinite_Pulses}, when $\pt < T_{\rm a}$, the optimal strategy for the (infinitely long) decreasing exponential pulse is to immediately transfer all excitations from $\ket{g}$ to $\ket{e}$ using an optical $\pi$ pulse. In this way, the pulse is retrieved as fast as possible with the shape of an exponential with the largest possible exponent. For this immediate transfer to happen, $\Omega(t)$ must be a delta function at $t=t_c=t_1$. 
This form of the drive is very challenging to capture in the optimization, which leads to the bad performance. 
This is true not only when $\pt < T_{\rm a}$, but also outside this regime. The rapid onset of the decreasing exponential still requires a delta-function drive shape, but with a smaller pulse area. This leads to the poor performance of the numerical method throughout the considered range of $\Gamma \ct $.
\begin{figure}
	\centering
	\includegraphics[width=\columnwidth]{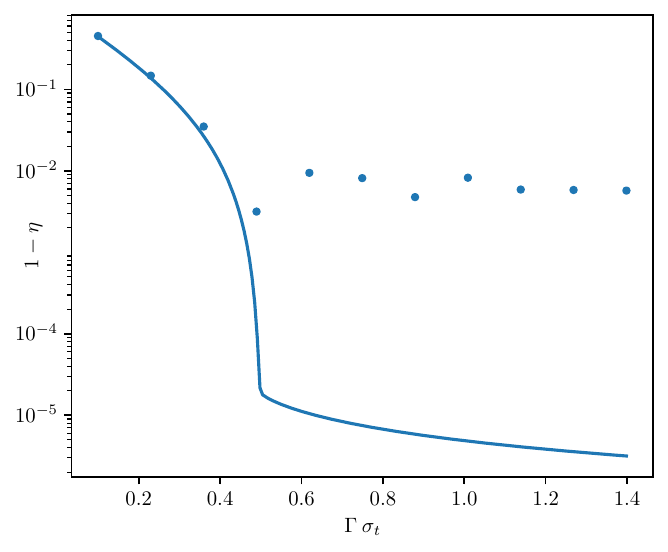}
	\caption{
Comparison between our protocol (solid line) and that obtained numerically via OCT (dots) for the decreasing exponential in Table~\ref{table:inf_pulses}. In this case, for the numerical optimization we split the time interval $t\in[t_1,t_2]$ into $n=150$ not evenly spaced time instants, with a higher density of points near the final time $t_2$ (recall that the numerical optimization is done for storage), and linearly interpolated them to obtain $\Omega(t)$. We present the inefficiencies for different values of $\Gamma \ct $. We set $\Delta = 0$, and truncate and renormalize $\phi$ such that $|t_2-t_1| = 2\sqrt{3} \pi \pt$ and $\int_{t_1}^{t_2}|\phi|^2(t) dt = 1$. We remark that, due to the finite interval, values of $\ct $ are not those in Table~\ref{table:inf_pulses}, but are calculated directly from Eq.~\eqref{eq:coh_time}.
}
	\label{fig:dec_exp_broad}
\end{figure}

In Fig.~\ref{fig:dec_exp_broad} we investigate whether it is possible, with our numerical method, to reach the infidelities obtained by our approach. Based on the analytically expected shape of $\Omega$, we vary the mesh employed in the numerical optimization. Specifically, for a fixed number of grid points, we greatly increased their density in the region where the derivative of $\Omega$ is higher. This improved the numerically evaluated efficiency, but still we were unable to reach performances comparable to our analytical approach. 
We  expect that, by increasing the number of grid points, it would be possible to improve the resulting efficiencies. We have, however, reached the limits of our computational resources, and further  analysis is beyond the scope of this work.
\bibliography{Bibliography}

\end{document}